\documentclass[epj]{svjour}

\usepackage{graphicx}	
\usepackage{amsmath}	
\usepackage{amssymb}	
\usepackage{array}
\usepackage{lineno}
\usepackage{multirow}
\usepackage{amsmath}
\usepackage{amsfonts}
\usepackage{amssymb}
\usepackage{bm}
\usepackage{bigstrut}
\usepackage{aas_macros}
\usepackage{placeins}
\usepackage{calc}

\begin{document}
\authorrunning{Sharma, Mukherjee, Jassal}
\titlerunning{Reconstruction of late time cosmology using PCA}
\title{Reconstruction of latetime cosmology using Principal Component Analysis}
\author{
Ranbir Sharma \inst{1}\thanks{E-mail: ranbirsharma0313@gmail.com}
Ankan Mukherjee \inst{2}\thanks{E-mail: ankan.ju@gmail.com}
H. K. Jassal \inst{1}\thanks{E-mail: hkjassal@iisermohali.ac.in}}

\institute{Indian Institute of Science Education and Research Mohali, SAS Nagar, Punjab 140306, India \and 
Centre for Theoretical Physics, Jamia Millia Islamia, New Delhi 110025, India.}

\abstract{
We reconstruct late-time cosmology 
using the technique of Principal Component Analysis (PCA).  
In particular,  we focus on the reconstruction of the dark energy
equation of state from two different observational data-sets,
Supernovae type Ia data, and  Hubble parameter data.  
The analysis is carried out in two different approaches.
The first one is a derived approach, where we reconstruct the observable
quantity using PCA and subsequently construct the equation of state parameter.  
The other approach is the direct reconstruction of the equation of state from the data.   
A combination of PCA algorithm and calculation of
correlation coefficients are used as prime tools of reconstruction.
We carry out the analysis with simulated data as well as with real data.   
The derived approach is found to be statistically preferable over the
direct approach. 
The reconstructed equation of state indicates a slowly varying equation of state of dark
energy.}

\PACS{{Cosmology,} {dark energy equation of state reconstruction,} {Principal Component Analysis,} {correlation coefficient}.}


\maketitle
\section{Introduction}
Cosmological parameters are now constrained to much better precision than before \cite{Aghanim:2018eyx}. 
This has been facilitated with significant improvements in
observational techniques and the accessibility of various
observational data.
Data in the last two decades has confirmed that the dynamics of the
Universe is dominated by a negative pressure component, known as dark
energy. 
Dark energy is understood to be the component driving  the observed 
accelerated expansion of the present Universe.
One of the prime endeavors of modern cosmological research is to
reveal the fundamental identity of dark energy, its exact nature and
evolution. 

It is not clear from the present observations whether dark energy
is a {\it cosmological constant}
\cite{Carroll1992,Carroll:2000fy,Turner:1998ex,Padmanabhan:2002ji}  or a
time-evolving entity \cite{Peebles:2002gy,Copeland:2006wr}.
The dark energy can be described by  equation of state parameter
$w=-P_{de} / \rho_{de}$,  where $\rho_{de}$ is the energy density and
$P_{de}$ is its pressure contribution.  
The form of the equation of state parameter $w$ of dark energy
depends on the theoretical scenario being considered.    
A constant value $w=-1$ corresponds to the  $\Lambda$CDM ({\it
  cosmological constant}  with {\it cold dark matter}) model, whereas
in case of time-evolving dark energy, the equation of state parameter can have different values \cite{Carroll1992,Padmanabhan:2002ji,Peebles:2002gy,Weinberg1989,Coble1997,Caldwell1998,Sahni2000,Ellis2003,Linder2008,Frieman2008,Albrecht2006,licia2008,nesseris2020}. 
We have little theoretical insight into these models, except for the
$\Lambda$CDM  model, which has a strong theoretical motivation.  
However, the standard  ($\Lambda$CDM) model faces the problem of
fine-tuning \cite{Carroll1992,Padmanabhan:2002ji,Weinberg1989,Coble1997}.
The observed value of the cosmological constant is found to be
many orders of magnitude smaller than the value calculated in quantum
field theories. 
Various alternative models have been proposed, which are based on
fluids, canonical and non-canonical scalar fields.
These models have the fine-tuning problem of their own, for instance,
scalar fields require  potentials which are specifically tailored to
match observations 
\cite{Ratra:1987rm,Linder:2006sv,Caldwell:2005tm,Linder:2007wa,Huterer:2006mv,Zlatev:1998tr,Copeland:1997et,PhysRevD.66.021301,Singh:2019bfd,Singh:2018izf,Bagla:2003prd,Tsujikawa:2013,Rajvanshi:2019wmw,Chevallier:2000qy}.

Since the approach to understanding dark energy is primarily
phenomenological, it is necessary to determine the model parameters to
constrain and rule out models that are not consistent with data.    
Likelihood analysis is the most commonly used technique in cosmological
parameter estimation and model fitting \cite{Singh:2019bfd,Singh:2018izf,Jassal:2009ya,Nesseris:2004prd,Nesseris:2005prd,Nesseris:2007,Archana:2018,Sangwan:2018zpz}.  
It is based on Bayesian statistical inference, where the posterior
probability distribution of a parameter is determined with a uniform
or variant prior function and the likelihood function.  
A combination of different data-sets, with likelihood regions
complementary to each other allows a very narrow range for
cosmological parameters.  
Thus the combination of various data-sets tightens the constraints on
the parameters.    
This {\it parametric reconstruction}
\cite{Chevallier:2000qy,Linder:2002et,Jassal:2004ej,Gong:2006gs,Mukherjee:2016eqj,2018PhRvD..98h3501V,DiValentino:2017zyq,licia2020_1,licia2020_2,licia:2013} assumes a functional form of the equation of
state parameter of dark energy.
However, it induces the possibility of bias in the parameter values
due to the assumption of the parametric form.   
One elegant way to constrain the equation of state parameter is by the differential 
age method of cosmic chronometer technique, which can 
be used to find out the evolution of the Universe without any inference from
a particular cosmological model \cite{licia2008,Jimenez:2001gg,Moresco:2018ApJ,Valent:2018}.

An alternative method is to reconstruct the evolution of cosmological
quantities in a  non-parametric fashion. 
Various statistical techniques have been  adopted for non-parametric
reconstruction of cosmological quantities \cite{Montiel2014,Gonz2016,Taylor2019,Porqueres2017,Diaz2019,Arjona2019prd}. 
One of the promising techniques in non-parametric approaches is
the Gaussian Process(GP)
\cite{Valent:2018,Sahlen:2005zw,Holsclaw2010am,Shafieloo:2012ht,Francesca:2019jcap,Rasmussen:2005,Bonilla2020}, where we can create
a multivariate Gaussian function with a  determined mean and
corresponding covariance function from any finite number of
collections of random variables.

To accomplish the reconstruction of cosmological quantities, we use
the Principal Component Analysis (PCA). 
A comparative study of different model-independent methods can be
found in \cite{Nesseris:2013PhRvD}. 
PCA is a multivariate analysis and is usually employed to predict the form of cosmological 
quantities in a model-independent, non-parametric manner \cite{Huterer2003prl,clarkson2010prl,Huterer2005,Zheng:2017ulu,Ishida2011is}. 
In \cite{Qin:2015eda} and \cite{Liu:2015yha}, different variants of 
PCA techniques have been adopted. 
Reference \cite{Qin:2015eda} uses an error model and then creates different set of 
simulated Hubble data to construct the covariance matrix while
\cite{Liu:2015yha}  use the weighted least square method and combines it
with PCA.

From the viewpoint of the application of PCA, there exist two distinct methodologies.
These methods differ mainly in the way the covariance matrix is calculated,
which is the first step of any PCA technique. 
One way to implement PCA is through the computation of Fisher matrix,
\cite{Nesseris:2013PhRvD,Huterer2003prl,clarkson2010prl,Huterer2005,Zheng:2017ulu,Ishida2011is,Crittenden:2005wj,Miranda:2018,Hart:2019,Hojjati:2012prd,2013JCAP...02..049N,Hart:2021kad}.  
One can bin the redshift range and assume a constant value for the quantity to be reconstructed 
in that redshift bin. 
These constant values are the initial parameters of the PCA.
Fisher Matrix quantifies the correlation and uncertainties of these parameters.
Therefore, by deriving these constants in different bins using PCA, we
can reproduce our targeted quantity in terms of redshift \cite{Ishida2011is}.
Alternatively, a polynomial expression for the dynamical quantity can be assumed.
In this case, the coefficients of the polynomial are the initial parameters for PCA, 
and the analysis gives the final values of these coefficients, which eventually
gives the dynamical quantity \cite{clarkson2010prl}.

PCA is independent of any prior biases and it is also helpful in comparing
the quality of different data-sets \cite{Crittenden:2005wj,yu2013prd}.
It is an application of linear algebra, which makes the linearly
correlated data points uncorrelated. 
The correlated data points of the data-set used in PCA
are transformed by rotating the axes, where the angle of rotation of these axes 
is such that linear-correlations between data-points are the smallest
compared to any other orientation.  
The new axes are the principal component(PC)s of the data points,
and these PCs are orthogonal to each other.
In terms of information in the data-set, PCA creates a hierarchy of priority
between these PCs.
The first PC contains information of the signal the most and hence
has the smallest dispersion of data-points about it.
The second PC contains less information than the first PC and
therefore provides a higher dispersion of the data-points as
compared to the first PC. 
Higher-order PCs have the least priority as these correspond to noise 
and we can drop them.
The reduction of dimensions is a distinctive feature of PCA. 
Therefore the final reconstructed curve in the lower dimension
corresponds predominantly to the signal of the data-set.  
The PCA method also differs from the regression algorithms, which can not distinguish 
between signal and the noise.
PCA can omit the features coming from the noise part and can pick the
actual trend of the data-points \cite{licia2010lnip,Sivia2006,Steinhardt2018}.
In our case, the data-points on which we apply PCA are created in the process of reconstruction, see sect(\ref{sec::algo}).


We assume polynomial expressions for the
observational quantities, namely the Hubble parameter $H(z)$, and the
distance modulus ($\mu(z)$). 
The polynomial form is then modified and predicted solely by PCA
reconstruction algorithm.  
The only assumption we make is that it is possible to expand the reconstructed quantity in a polynomial. 
The dark energy equation of state is constructed in the subsequent steps after the reconstruction of the Hubble parameter and distance modulus.


The two-step process of reconstruction is termed as the {\it derived approach}.
Further, in a {\it direct approach}, we apply the PCA technique 
to the polynomial expression of the equation of state parameter $w(z)$. 
For the application of our methodology, we require a tabulated
  dataset of the observational quantity, which can schematically be
  represented as, \textit{(independent variable)--(dependent
    variable)--(error bar of dependent variable)}.  
In the direct approach, described below, we do not have the tabulated data-set of $w(z)$ and we derive the functional form of $w(z)$ from the tabulated data-set of $z - H(z)$ and $z - \mu(z)$.
In the case of the derived approach, PCA first reconstructs the functional
form of $H(z)$ and $\mu(z)$ from the Hubble parameter and
Supernovae data-set respectively without any cosmological model.
In the absence of cosmological model, $H(z)$ and $\mu(z)$ have no any parameter dependencies.
However to construct the final form of equation of state of dark energy, $w(z)$ we need $\Omega_m$ as input.
Application of PCA on polynomial expressions to reconstruct
cosmological quantities have been employed earlier in
\cite{Ishida2011is,Qin:2015eda,Liu:2015yha}. 


For a further check on the robustness of our results; we compute the
correlations of the coefficients, present in the polynomial
expressions of the reconstructed quantities. 
We apply PCA in the coefficient space, which is created by the 
polynomial.
The  PCA rearranges the correlation of these coefficients.  
Calculation of correlation-coefficients help to identify the linear
and non-linear correlations of the components in the reconstructed
quantities.   
We show that using the correlation coefficients calculation, 
we can restrict the allowed terms in the polynomial. 

This paper is structured as follows. In sect(\ref{sec::algo}) 
we describe the reconstruction algorithm and
the use of correlation tests in those reconstructions. 
In sect(\ref{sec::reconstruction}), we describe the two approaches of reconstruction we follow.
We present our results in sect(\ref{sec::results}).
In sect(\ref{sec::conclusion}), we conclude by summarizing
the results.   

\section{Methodology} \label{sec::algo}
We begin with an initial basis, $g_i = f(x)^{i - 1}$, where
$i=1,2,....,N$ through which we can express the quantity to be
reconstructed as,
\begin{equation} \label{polynomial}
\xi(x) = \sum_{i=1}^N b_if(x)^{(i-1)} 
\end{equation}

The initial basis can be written in matrix form as, $\mathbf{G} =
(f_1(x),f_2(x),...,f_N(x))$.  
Coefficients ${b_i}$ create a {\it coefficient space} of dimension $N$.
Each point in the coefficient space gives a realization of $\xi(x)$,
including a constant $\xi(x)$. 
PCA modifies the coefficient space and chooses a single realization.
One can consider different kinds of functions as well as different combinations of polynomials as  initial bases.
Choosing one initial basis function over the other is done by correlation coefficient calculation described below.

The Pearson correlation coefficient for two parameters $A$ and $B$ is
given by,   
\begin{equation}
\rho = \frac{Cov(A,B)}{\sigma_A \sigma_B}
\end{equation} 
where $\rho \in [-1,1]$.
For linearly uncorrelated variables, the correlation coefficient, $\rho=0$.  
An exact correlation is identified by $\rho = -1$ or $\rho = +1$.
The Spearman rank coefficient is, in turn, the Pearson correlation 
coefficient of the ranks of the parameters; rank being the value 
assigned to a set of objects and it determines
the relation of every object in the set with the rest of them. 
We  mark the highest numeric value of a variable $A$ as ranked
1, the second-highest numeric value of the variable as ranked 2 and so
on.
A similar ranking is done for the ranks of the parameter $B$. 

To obtain the  coefficients for the correlation analysis, we divide
the parameter space into $n$ patches. 
We therefore have $n$ values associated with one coefficient of the polynomial; 
this is the number of columns of the coefficients matrix $\mathbf{Y}$,
(eq(\ref{eqn::coefficient_matrix})).  
After ranking all the values of $A$ and $B$, we obtain the 
table for the ranks of $A$ and $B$. 
We then proceed to compute the  Spearman correlation coefficient($r$),
which is the Pearson Correlation coefficient of rank of $A$ and $B$. 
Like in the case of the Pearson Correlation coefficients, $r \in [-1,1]$. 
Computing Kendall correlation coefficient($\tau$) gives a prescription
to calculate the total number of concordant and dis-concordant pairs
from the values of the variables $A$ and $B$  
\cite{M_G_Kendall_1938}.

If we pick two pairs of points from the table of $A$ and $B$,
say $(a_i,b_i)$ and $(a_j,b_j)$,  for $i \neq j$ if $a_i > a_j$ when $b_i
> b_j$ or if $a_i < a_j$  when $b_i < b_j$; then that pair of points
are said to be in concordance with each other.  
On the other hand, for $i \neq j$, $a_i > a_j$ when $b_i < b_j$ or if $a_i <
a_j$ when $b_i > b_j$, then these two pairs are called to be in
dis-concordance with each other. 
Every concordant pair is scored $+1$ and every dis-concordant
pair is scored  $-1$.
The Kendall correlation coefficients are defined as,
\begin{equation}
\tau = \frac{\text{actual score}}{\text{maximum possible score}}
\end{equation}
\begin{displaymath}
\text{maximum possible score} =\frac{n(n-1)}{2}
\end{displaymath}
Again, if $N_{cp}$ is the number of concordance pair and $N_{dp}$ is
the number of dis-concordance pair
\begin{displaymath}
\text{actual score} = N_{cp} - N_{dp} 
\end{displaymath}
Hence the expression of $\tau$ is,
\begin{equation}
\tau = \frac{N_{cp} - N_{dp}}{n(n-1)/2}
\end{equation}
where, $\tau \in [-1,1]$.

We perform the correlation coefficients calculation twice.  
The first time is to select the number of terms in the initial polynomial
$N$(eq(\ref{polynomial})), 
and second time to select the number of terms in the final polynomial,
(eq(\ref{polynomial_second})).  
We select that value of $N$ for
which the Pearson Correlation Coefficient is higher than the Spearman
and Kendall Correlation coefficients.  
We use the R-package for statistical computing to calculate the
correlation coefficients \cite{R:manual}.

Values of linear and non-linear correlations depend on the quantity we want to reconstruct. 
They are also sensitive towards the data-set that we use in reconstruction.
For instance, reconstruction of a fast varying function, 
which has non-zero higher order derivatives will introduce more non-linear contributions to the correlation of $b_i$ than linear 
contributions, and we need a greater set of initial basis, which means a higher
value of $N$.      
We take different values of $N$ and check the linear and
non-linear correlation coefficients to fix the value of $N$.  
Though a large value would help, we can not, however, fix $N$ to any arbitrarily 
significant number as it makes the analysis computationally expensive.

To compute the covariance matrix, we define the coefficient
matrix ($\mathbf{Y}$) by selecting different patches from the coefficient space,
\begin{equation}
\mathbf{Y} = \mathbf{\bm{[}\bm{[}b_i^j\bm{]}\bm{]}}_{i=1,n}^{j=1,N}
\label{eqn::coefficient_matrix}
\end{equation}
where $n$ is the number of patches that we have taken into account and
$N$ is the total number of initial basis defined in
eq(\ref{polynomial});
therefore, Y is a matrix of dimension $N \times n$ and
$b_{n}^{(N)}$ being  the value of Nth
coefficient in $n$th patch.

In the present analysis, we have taken $n$ to be the order of
$10^3$.  
We estimate the best-fit values of the coefficients at each patch by $\chi^2$
minimization, where $\chi^2$ is defined as 
\begin{equation}
\chi^2=\sum_{j=1}^k\frac{(\xi(x)_{data}-\xi(\{b_i\},x))^2}{\sigma_j^2}
\label{ChiSquare}
\end{equation}
$k$ is the total number of points in the data-sets.
If the observational data-set have significant non-diagonal elements in the 
data covariance matrix($C_{data}$), we have to incorporate $C_{data}$ in eq(\ref{ChiSquare}), 
as $\chi^2 = \Delta^T \mathcal{C}_{data}^{-1} \Delta$, where
$\Delta = \xi(x)_{data}-\xi(\{b_i\},x)$ and $x$, $\xi(x)_{data}$ vary for each data-points.
Calculation of $\chi^2$ in all the patches gives us the variation of
the $N$ coefficients and finally gives $n$ number of points in the
coefficient space, over which we apply PCA.
We calculate the covariance matrix and correlations of the
coefficients for these $n$ points. 
In this analysis, each patch contains the origin of the
multi-dimensional coefficient space.   

The covariance matrix of the coefficients, $\mathbf{C}$ is written as,
\begin{equation*}
\mathbf{C} =\frac{1}{n}\mathbf{Y}\mathbf{Y}^T
\end{equation*}
Eigenvector matrix, $\mathcal{E}$ of this covariance matrix will
rotate the initial basis of the coefficient space to a position where
the patch-points will be uncorrelated.  
We organize the eigenvectors in the eigenvector matrix $\mathcal{E}$ 
in the increasing order of eigenvalues. 
Eigenvalues of the Covariance matrix quantifies the error associated with each principal component \cite{Huterer2003prl,Ishida2011is,2019Gortler,Louis1953}.

If $\mathbf{U} = (u_1(x),u_2(x),...,u_N(x))$ the final basis is given
by,  
\begin{equation}
\mathbf{U} = \mathbf{G}\mathcal{E}   
\end{equation}
The final reconstructed form of $\xi(x)$ is, 
\begin{equation}
    \xi(x) = \sum_{i=1}^M \beta_i u_i(x)
\label{polynomial_second}
\end{equation}
where $M \leq N$ and the $\beta_i$s are the uncorrelated coefficients
associated with the final basis.  
The coefficients $\beta_i$ are  re-calculated by using $\chi^2$
minimization, for the same $n$ patches considered earlier to create
initial  coefficient matrix (\ref{eqn::coefficient_matrix}). 
The value of $M$ can be determined by correlation coefficient
calculation discussed above.     
As PCA only breaks the linear correlation,  we select $M$ for which
PCA is able to break Pearson correlation coefficient to the largest
extent. 
Eigenvalues of the covariance matrix ($e_1, e_2, e_3,...,e_M$) indicate how well PCA can pick 
the best patch-point in the coefficient space. 
We look for the lowest number of final basis which can break the linear correlation to the greatest extent. 
We can start with smaller value of $N$ which eventually influence the value of $M$, but it will pose 
the risk of losing essential features from PCA data-set.


\section{Reconstruction of dark energy equation of state} \label{sec::reconstruction} 
   
The Hubble parameter ($H(z)$) for a spatially flat Universe, composed
of dark energy and non-relativistic matter is given by,    
\begin{equation} \label{eq:FRW}
H^2(z) = H_0^2[\Omega_m (1+z)^3 + \Omega_x e^{3 \int^z_0 \frac{1+w(z')}{1+z'}dz'}]
\end{equation}
Here we have assumed that the contributions to the energy density is only due to 
the non-relativistic dark matter and  dark energy.
The density parameters for non-relativistic matter and dark energy are
given by $\Omega_m$ and $\Omega_x$.
The quantity $H_0$ denotes the present-day value of the  Hubble parameter,
namely the Hubble constant and $w(z)$ is the dark energy equation of state parameter ($w(z)$).    

We assume no interaction between matter and
dark energy in the present analysis.
In the following subsections, we discuss the derived and direct approach to 
reconstruct $w(z)$ using PCA.

\subsection{Derived Approach} \label{sec::derived}
The derived approach is a two-step process in the reconstruction of dark energy 
Equation of State(EoS).  
In the  first step, we reconstruct the observable, namely the Hubble
parameter using the $H(z)$ data and 
the distance modulus $(\mu(z))$ using the type Ia supernova data. 
Subsequently, we reconstruct $w(z)$ as a derived quantity from these two different
physical quantities. 
Similar sequence of reconstructions have already been discussed in
\cite{Huterer2003prl,Huterer2005,Ishida2011is,yu2013prd}. 
We follow the approach mentioned in sect(\ref{sec::algo}) to
reconstruct the curve of $H(z)$ and $\mu(z)$.   
Differentiating eq(\ref{eq:FRW}) with redshift $z$  as the argument and rearranging the terms we
can express $w(z)$ as,  

\begin{equation}
w(z) = \frac{3H^2 - 2(1+z)HH'}{3H_0^2(1+z)^3 \Omega_M - 3H^2}
\label{eq:wz}
\end{equation}
Here, $H'$ is the derivative of Hubble parameter with respect to redshift $z$.
Since $w(z)$ is related to $H(z)$ through eq(\ref{eq:wz}) by the
zeroth and the first order differentiation of $H(z)$, the small
difference in the actual and  the reconstructed curve of $H(z)$ is
amplified by the $H'$ term.   
This process of amplification of the deviation from actual nature
becomes more severe with subsequent higher-order differentiation of
the reconstructed quantity.    

The luminosity distance $d_L(z)$ is given by,
\begin{equation} \label{eq:d_L}
d_L(z) = \frac{c}{H_0}(1+z)\int_0^z d_H(z')dz'
\end{equation}

where $d_H$, in terms of eq(\ref{eq:FRW}) is, 
\begin{equation} \label{eq:d_H}
d_H(z) = \left(\Omega_m (1+z)^3 + \Omega_x e^{{3\int_0^z \frac{(1+w(z'))dz'}{(1+z')}}}\right)^{-1/2}
\end{equation}

and is related to the distance modulus as  
\begin{equation} \label{eq:mu}
\mu(z) = 5 \log{\left(\frac{d_L}{1Mpc}\right)} + 25
\end{equation}
which is the dependent-variable in the type Ia supernovae data.   
From PCA, we determine the form of $\mu(z)$ directly from data and then 
from eq(\ref{eq:d_L}) and eq(\ref{eq:d_H}) find the expression of $d_L$.
From eq(\ref{eq:mu}), we trace back to eq(\ref{eq:d_H})
and find an expression which gives the $w(z)$ in terms of
the distance modulus. 

Since $D(z) = (H_0 / c) (1+z)^{-1} d_L(z) $, the equation of state
parameter is given by
\begin{equation} \label{eq:EoS_SNIa}  
w(z) = \frac{2(1+z)D'' +3D'}{3D'^3 \Omega_m (1+z)^3 -3D'}
\end{equation}
The second order derivative in eq(\ref{eq:EoS_SNIa}) makes the
reconstruction of $w(z)$ through that of distance modulus
unstable.
For instance, if the reconstruction fails to pick some of the
minute difference in the observational curve, then that difference
will be amplified twice in the final calculation of the EoS.
Therefore, the reconstruction of $\mu(z)$ should be more accurate in picking up
approximately  all the features of $w(z)$ which may be  hidden
within the supernovae data
\cite{Jimenez:2001gg,clarkson2010prl,Ma_and_Zhang_2011rs,pan_and_alam_2010pa}.

In the reconstruction of $H(z)$ and $\mu(z)$, we begin with 
polynomial expansions in terms of the different variables $z$, $a$ and
$(1-a)$ where $z$ is the red-shift and $a$ is the scale factor. 
The analysis is carried out with seven terms in initial basis, which
means creating a coefficient space of $N=7$ dimensions
(eq(\ref{polynomial})).  
We test our algorithm on a simulated data-set of $\Lambda$CDM cosmology.
We generate the data-points of $H(z)$ and $\mu(z)$ for $w(z) = -1$ and the values of cosmological parameters
$\Omega_m$ and $H_0$ are fixed at Planck 2018 values \cite{Aghanim:2018eyx}. 
Fig(\ref{fig::EoS_lcdm}) and table \ref{Table::H_constant_fidu} show that our algorithm can predict the  simulated data.


\subsection{Direct Approach} \label{sec::direct}

For the direct reconstruction approach, we begin with a polynomial
form of $w(z)$  itself.
In eq(\ref{eq:FRW}), the quantity $w(z)$ is in the exponent, and
considering a polynomial form for  $w(z)$ implies addition of some
non-linear components to  our linear analysis in coefficient space.  
Again we fix the dimension of the initial basis $N$ by computing  the 
correlation coefficients.
Here we need to balance available computational power as
well as   the non-linearity we introduce, with the accuracy we demand
to choose the value  of $N$.

In this case too the independent variables are taken to be $(1 - a)$, $a$ and $z$.  
Here we introduce non-linear terms in the initial coefficients of PCA; 
here correlation coefficients calculation is not of assistance as in
the case of derived approach \ref{sec::derived} to select $M$.
Due to the risk of compromising different features of the data-set and
also due to the  complex dynamics of correlation coefficients, we do
not reduce any terms in the  final basis ($N = M$) of the direct approach.

To test the effectiveness of both the approaches described above, we first
work with simulated data-sets for specific models.  
We create simulated data-points for $w=-1$ ($\Lambda$CDM) and $w(z)=-\tanh{(1/z)}$ \cite{Qin:2015eda} where the values of $\Omega_m$ and $H_0$ are fixed at Planck 2018 values \cite{Aghanim:2018eyx}.  
We use eq(\ref{eq:FRW}) to calculate $H(z)$ at the same redshift
value as in the real Hubble parameter vs redshift data-set
\cite{ohd1,ohd2,ohd3,ohd4,ohd5}. 
Similarly, the distance modulus data points are simulated using equations 
(\ref{eq:d_L}),(\ref{eq:d_H}) and (\ref{eq:mu}). 
Here we evaluate 
distance modulus $\mu(z)$ at the same redshift values  as are there 
in type Ia supernovae (SNe) Pantheon data-set \cite{panth_snIa}.
We later utilize the observational measurements of
Hubble parameter at different redshift \cite{ohd1,ohd2,ohd3,ohd4,ohd5}
and the distance modulus measurement of type Ia supernovae (SNe) data
\cite{panth_snIa}.


\section{Results}\label{sec::results}

\begin{figure*}
\begin{center}
 \includegraphics[width=5.5cm, height=3.5cm]{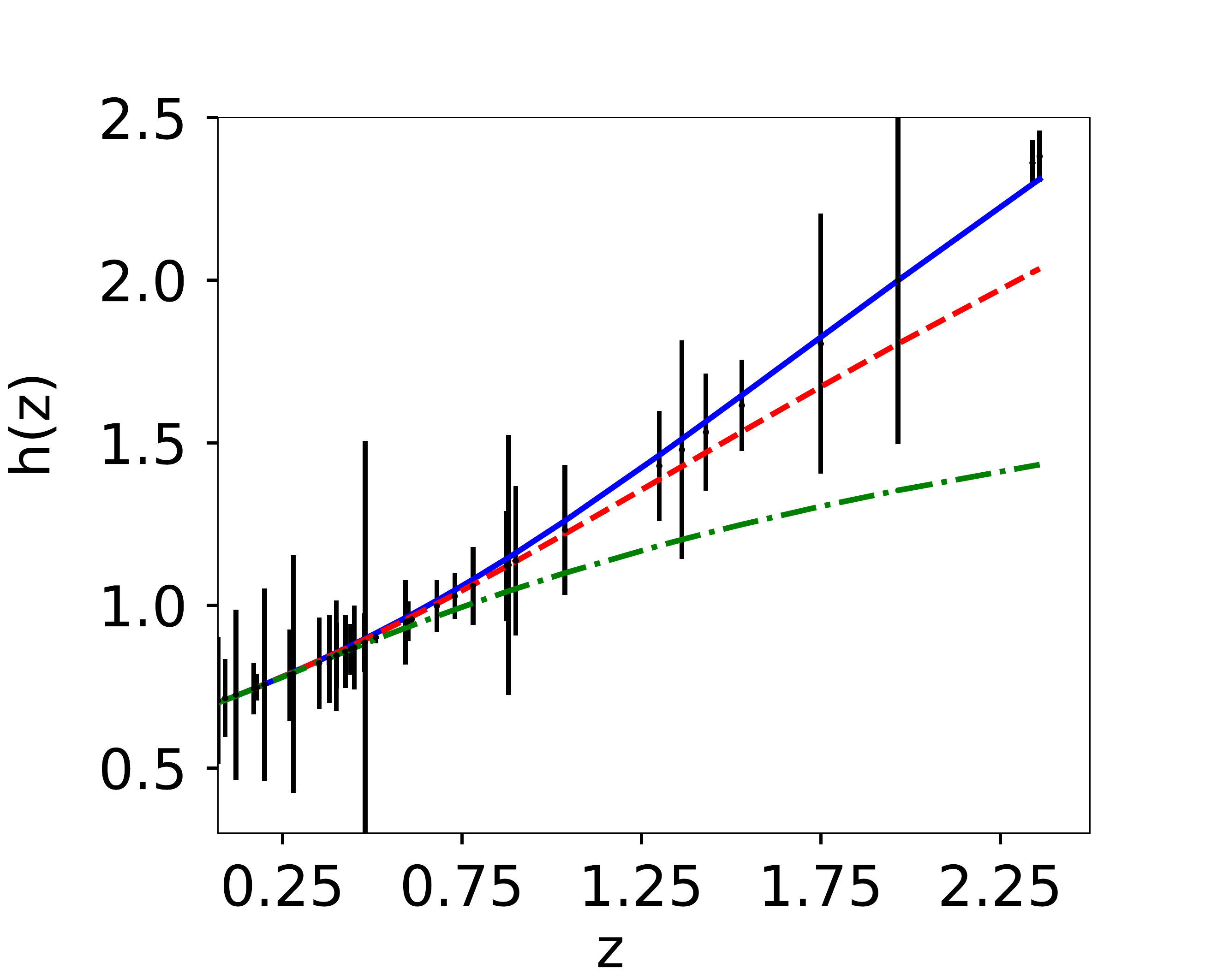}
 \includegraphics[width=5.5cm, height=3.5cm]{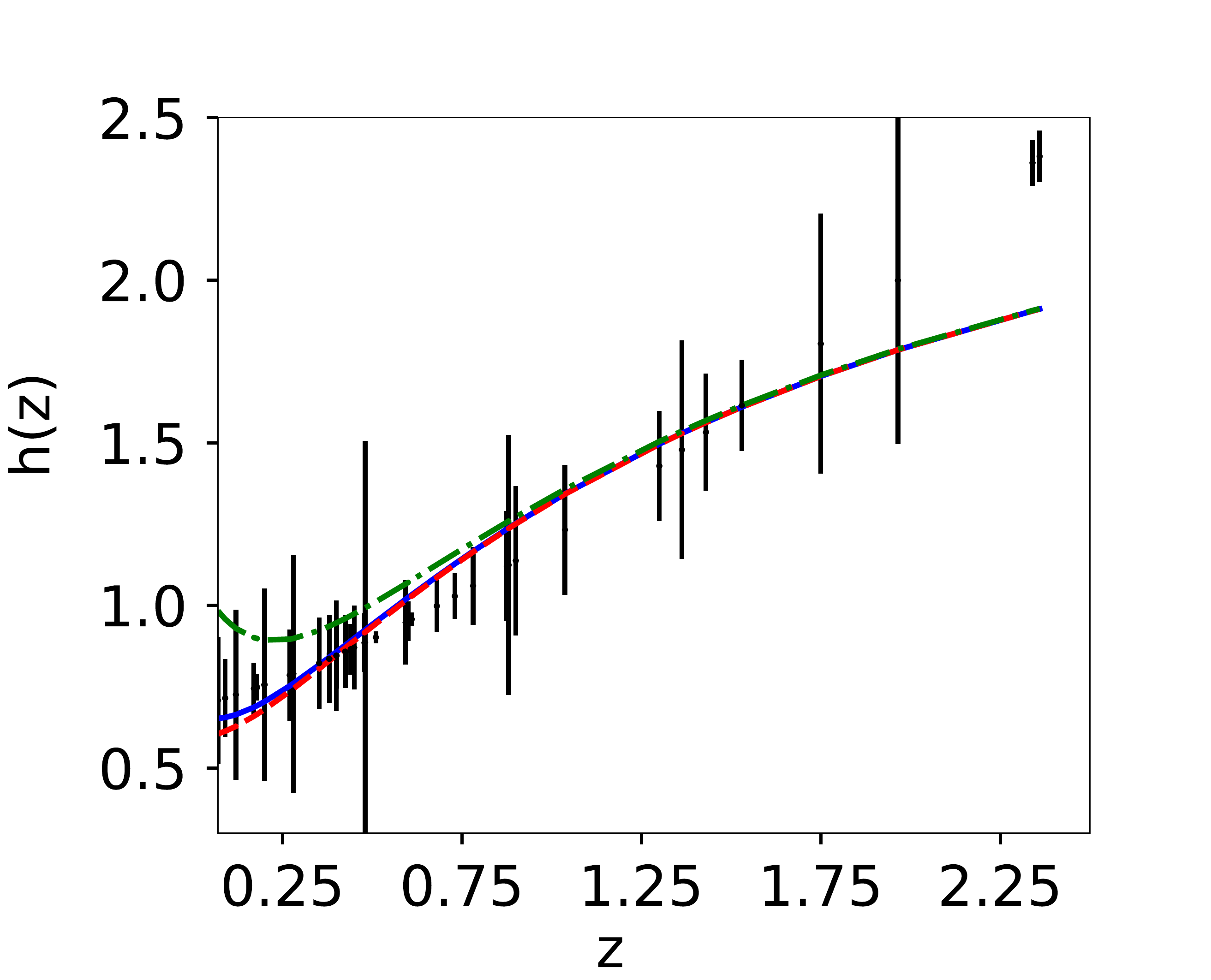} 
 \includegraphics[width=5.5cm, height=3.5cm]{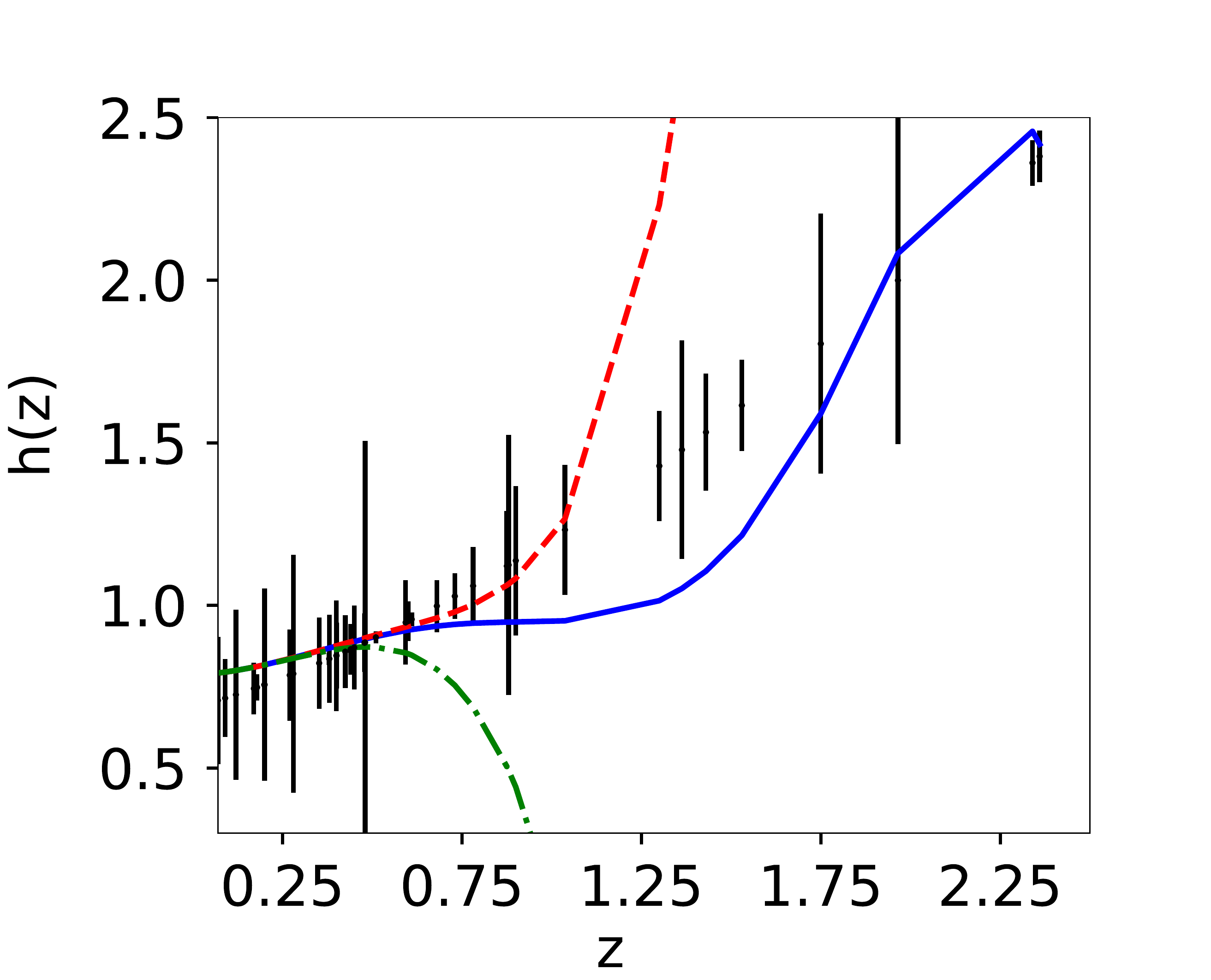}
 \includegraphics[width=5.5cm, height=3.5cm]{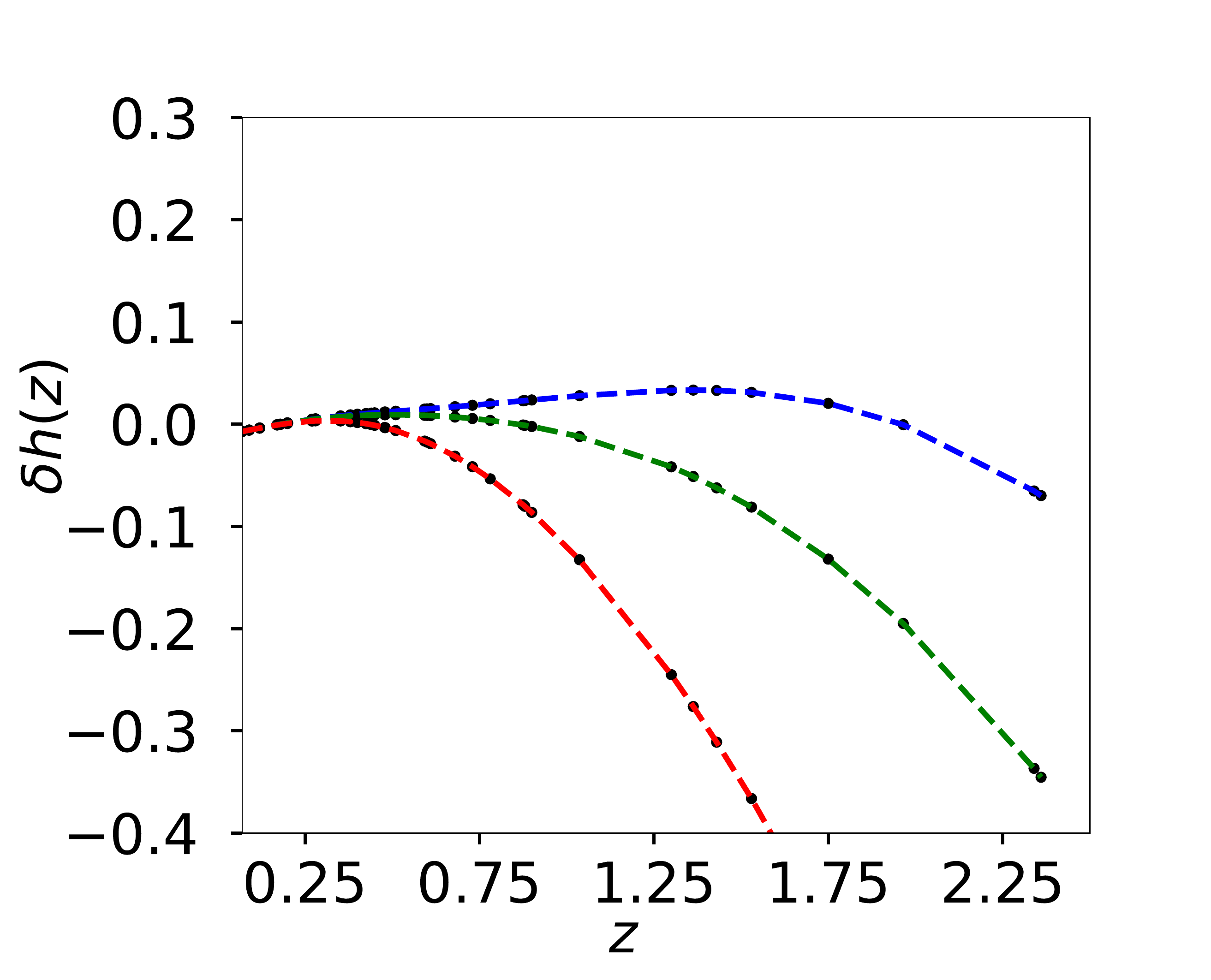}
 \includegraphics[width=5.5cm, height=3.5cm]{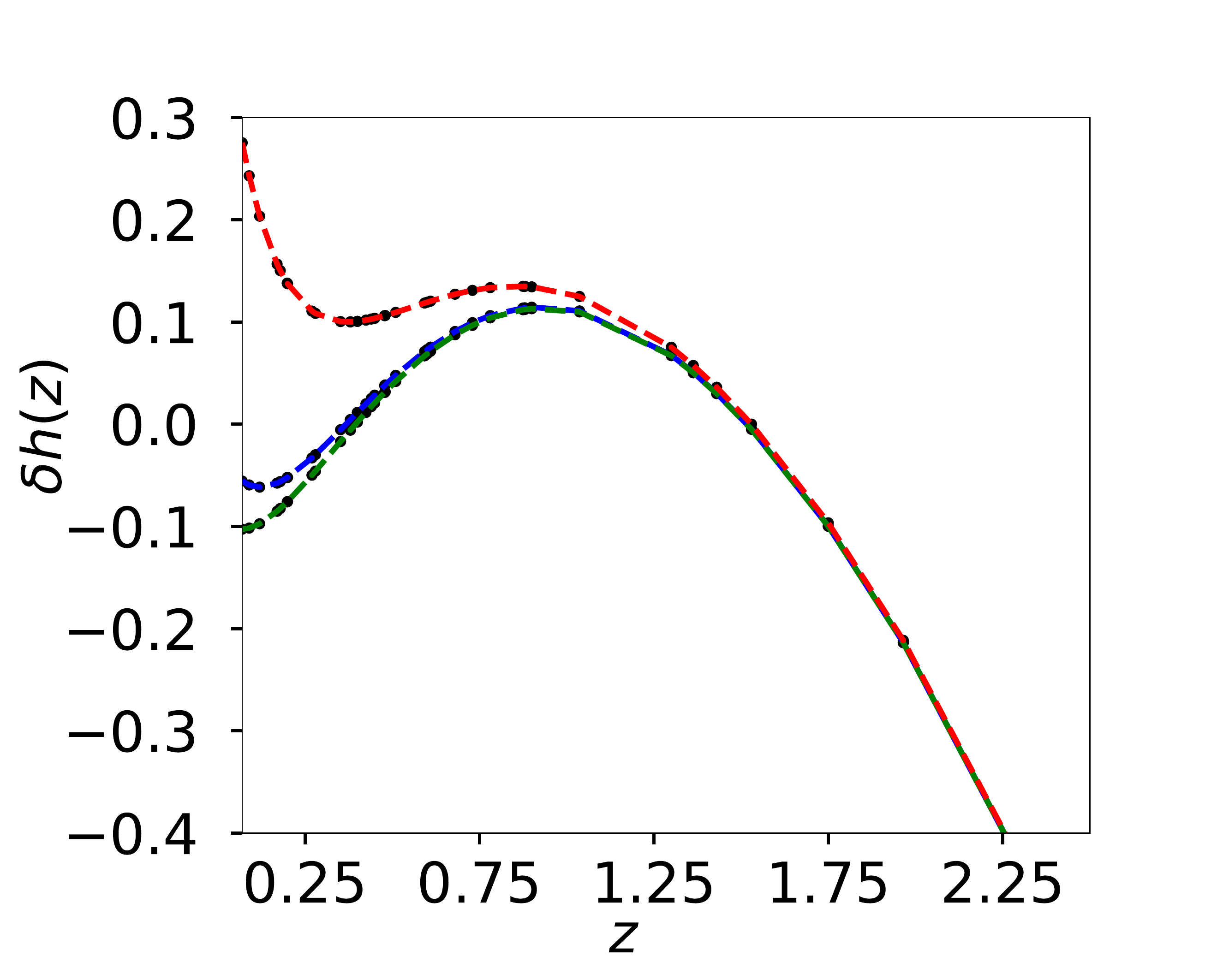}
 \includegraphics[width=5.5cm, height=3.5cm]{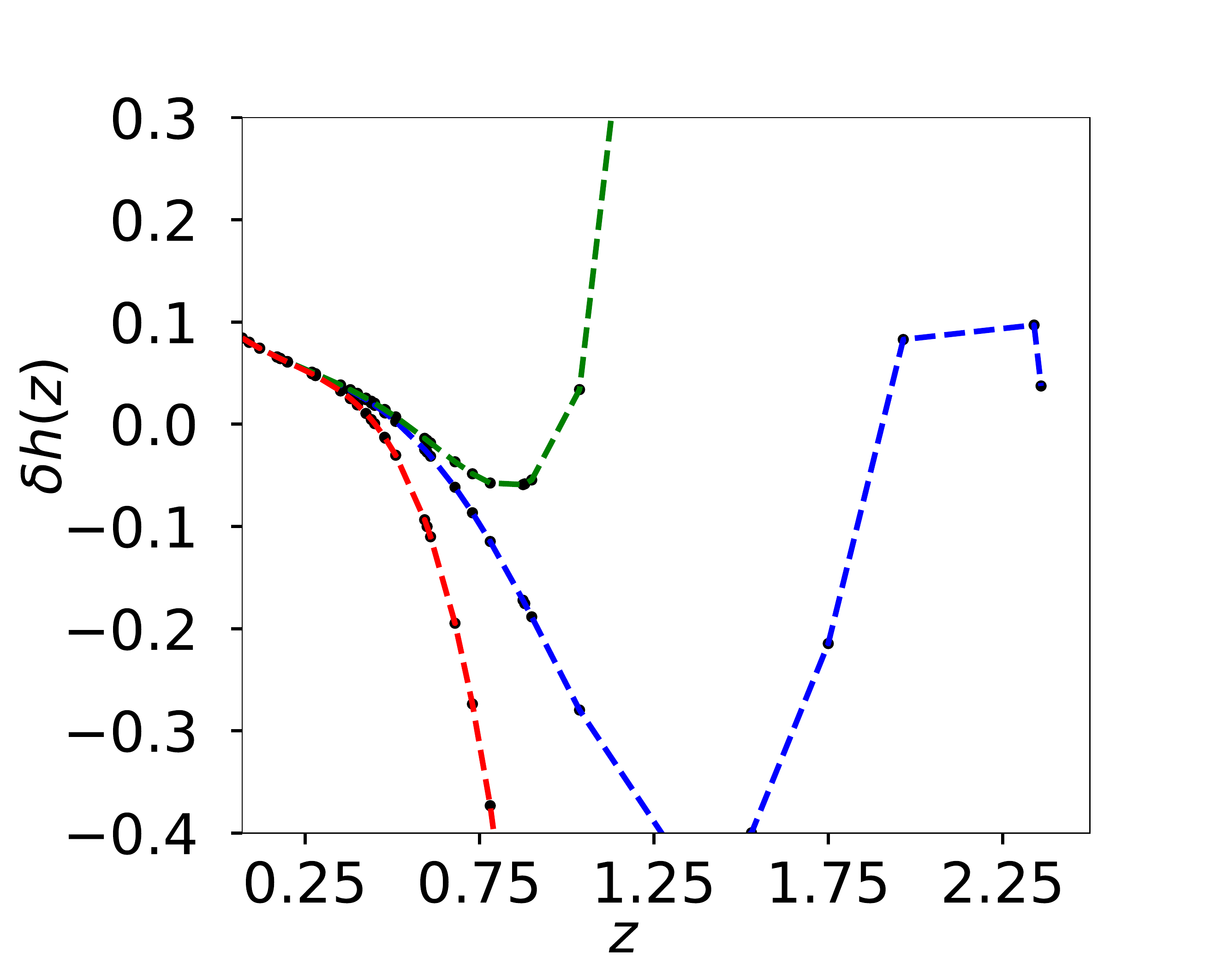}
 \includegraphics[width=5.5cm, height=3.5cm]{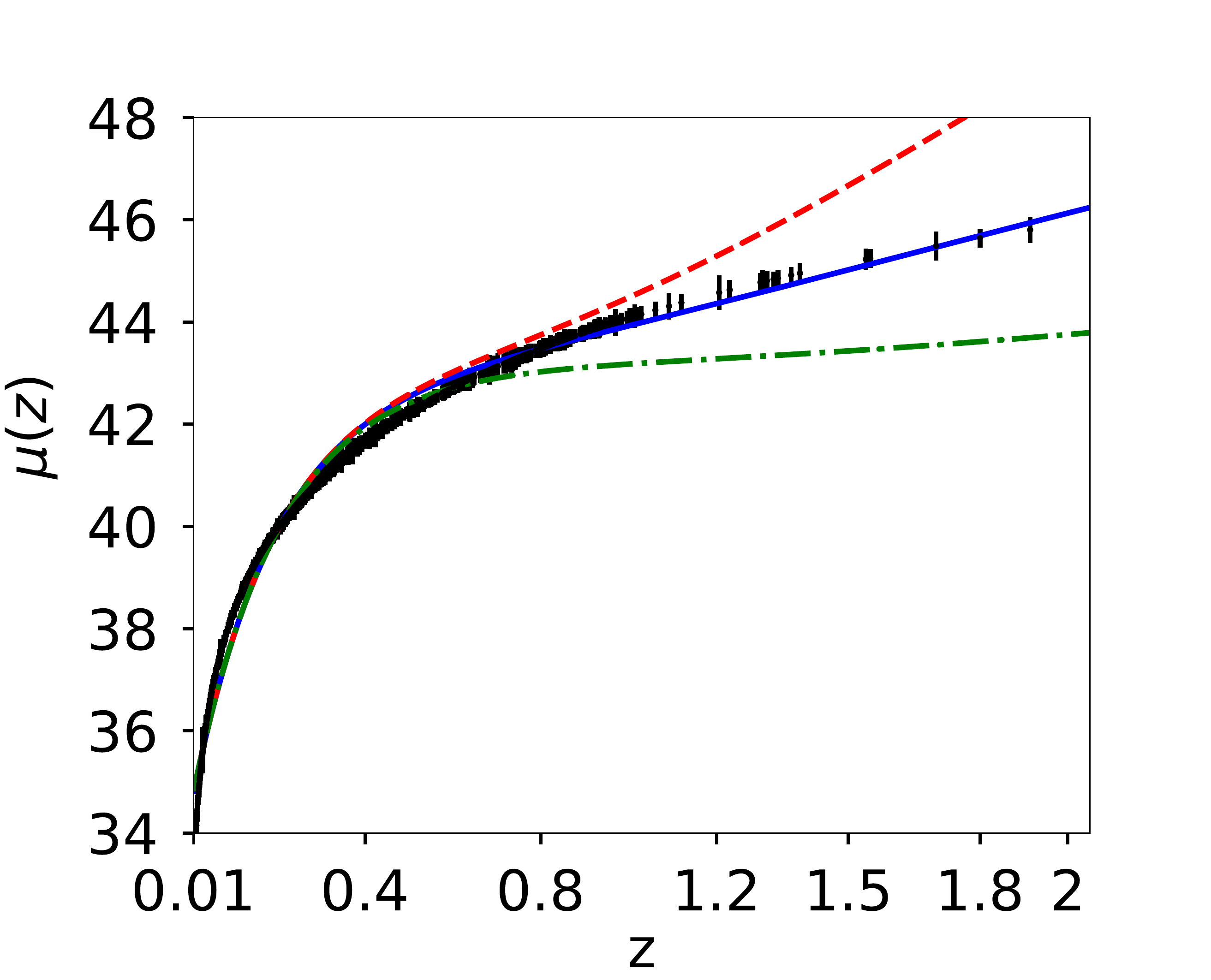}
 \includegraphics[width=5.5cm, height=3.5cm]{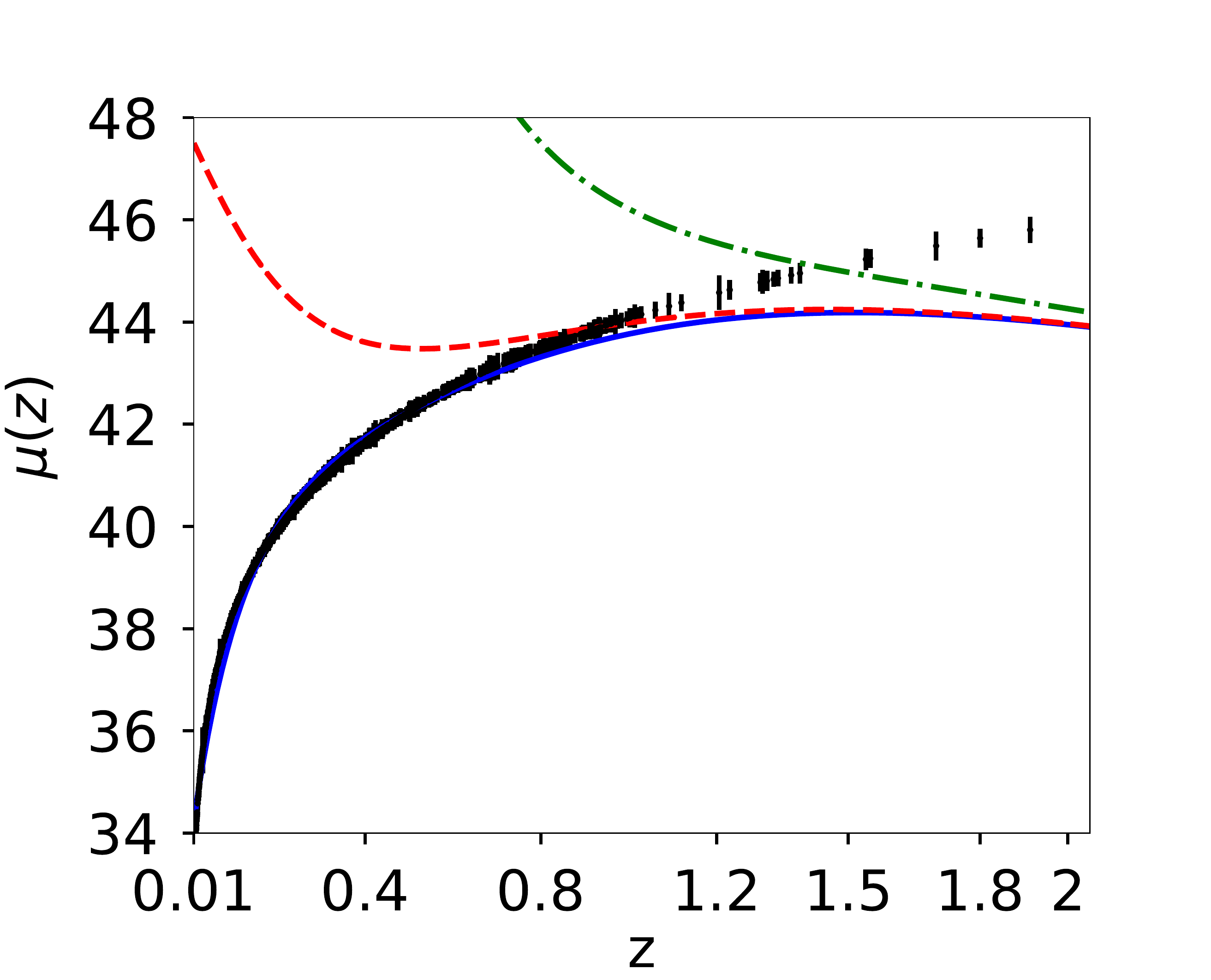}
 \includegraphics[width=5.5cm, height=3.5cm]{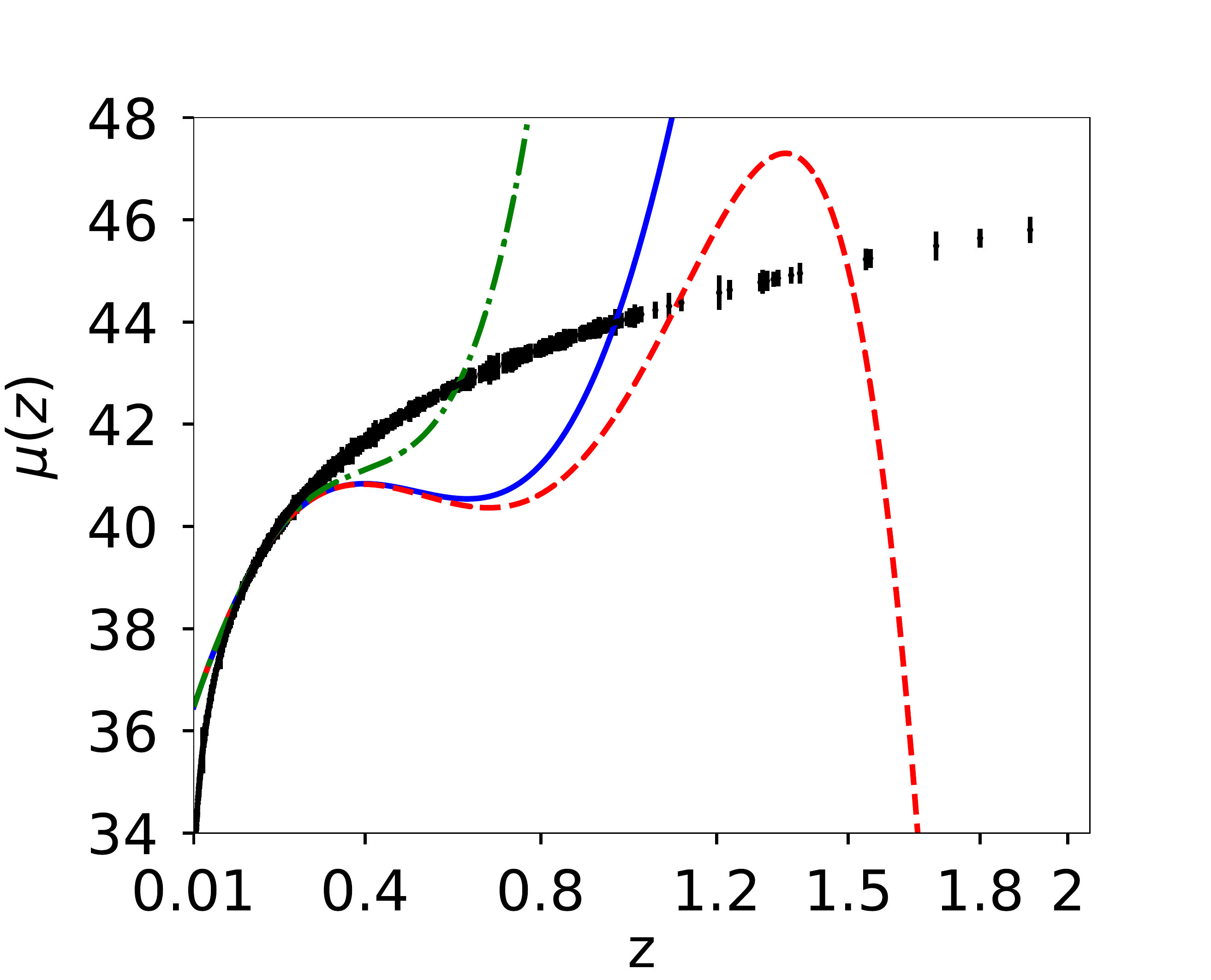}
 \includegraphics[width=5.5cm, height=3.5cm]{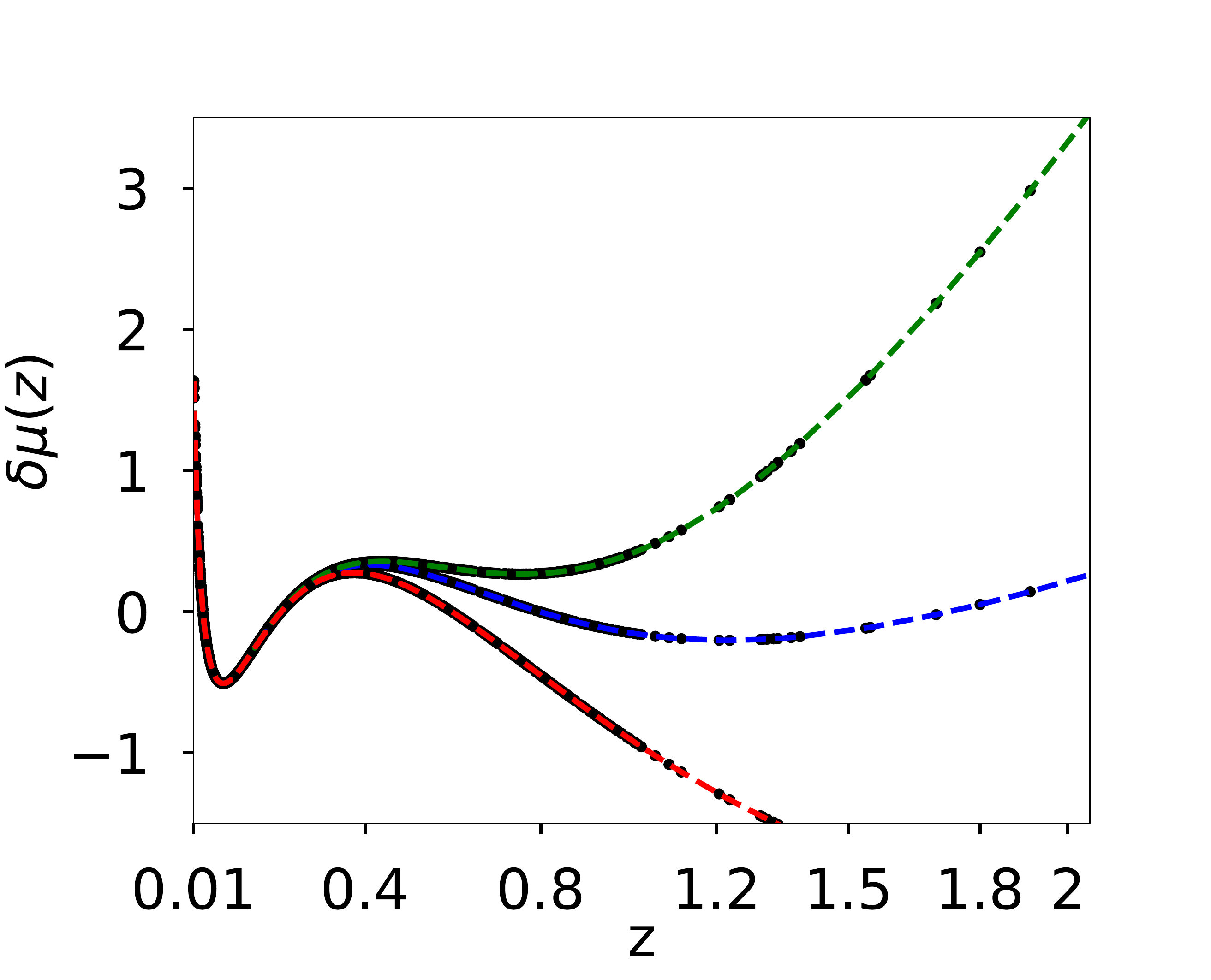}
 \includegraphics[width=5.5cm, height=3.5cm]{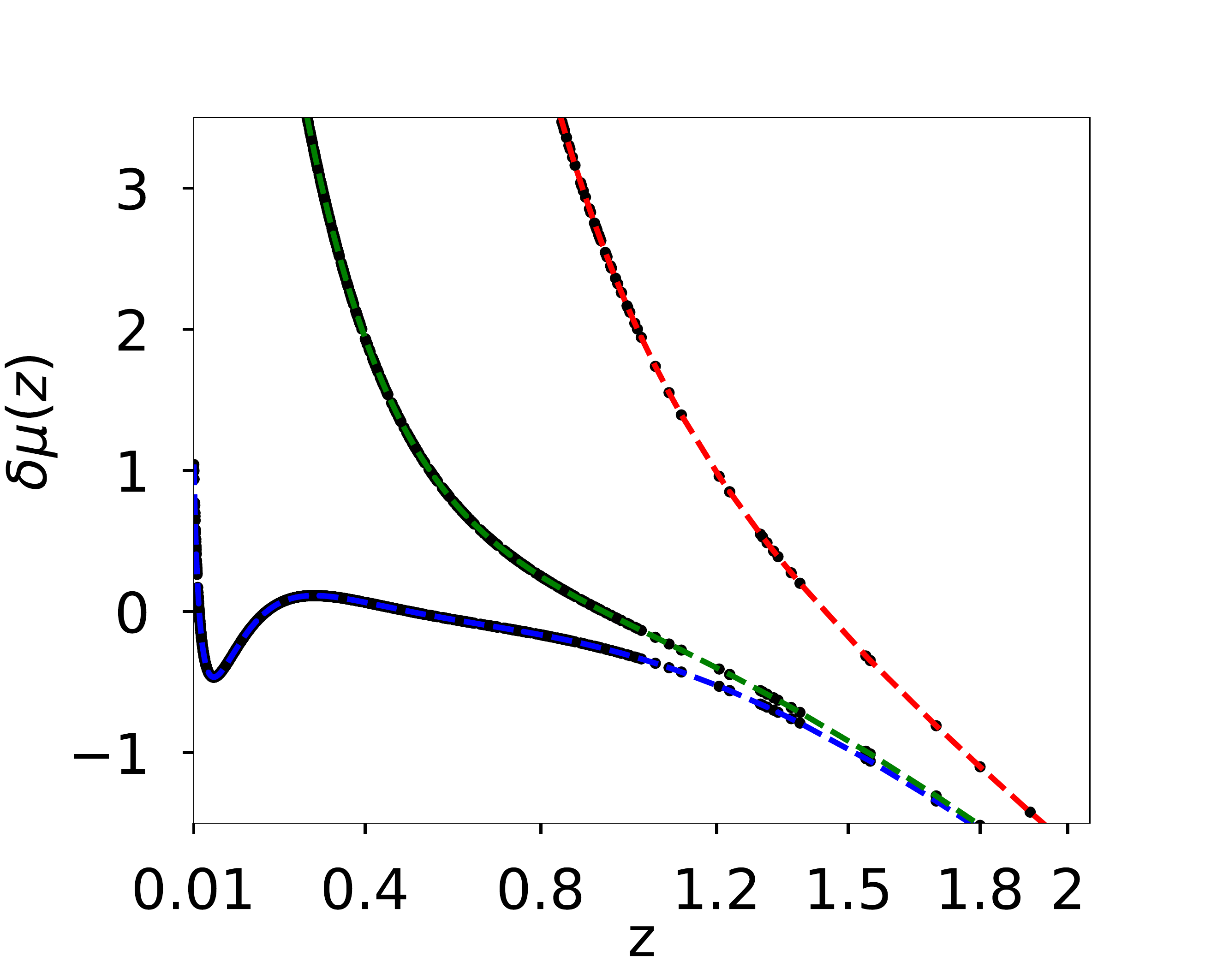}
 \includegraphics[width=5.5cm, height=3.5cm]{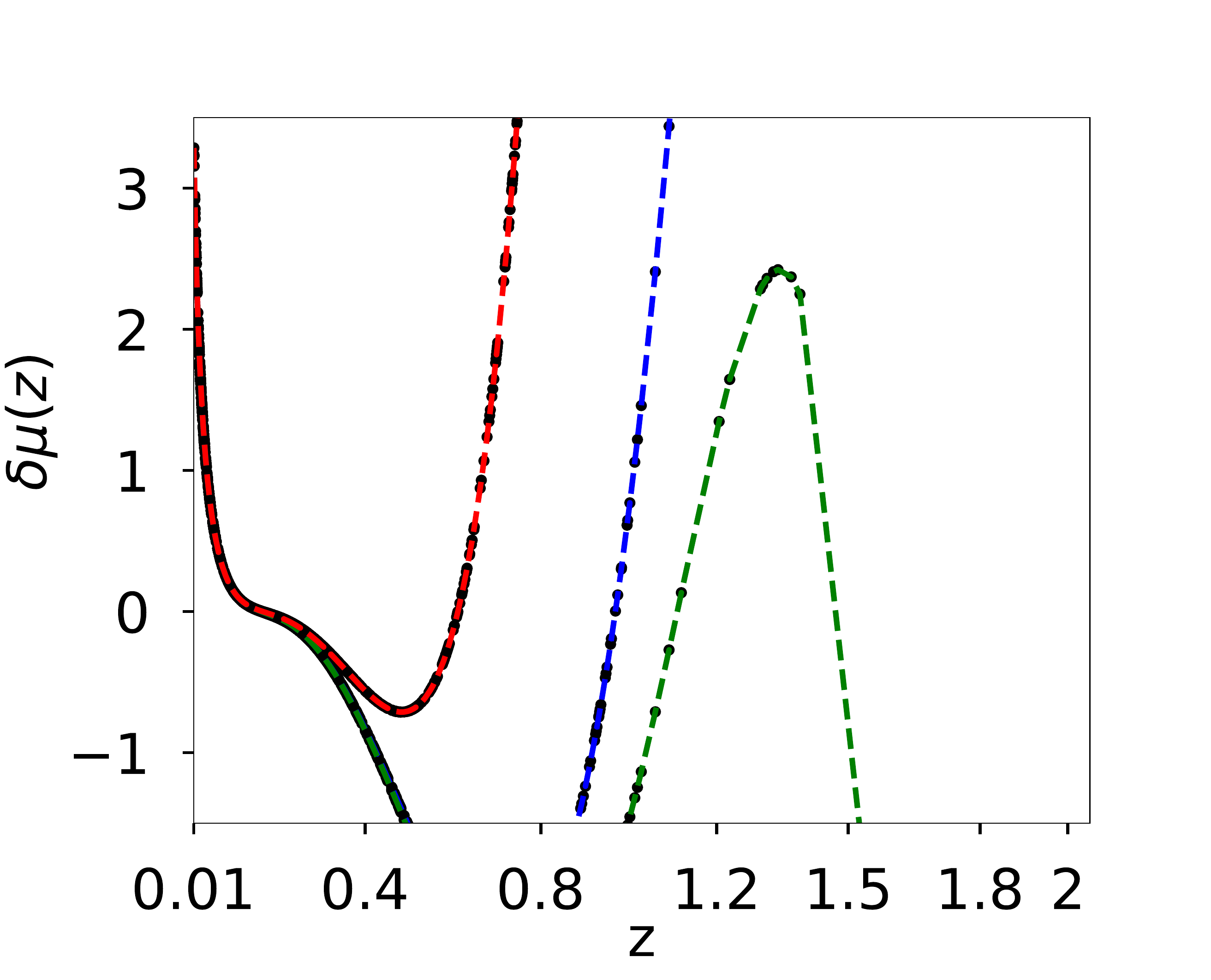}
\end{center}
\caption{The plots in the figure show the reconstructed reduced Hubble parameter $h(z)$ and
distance modulus $\mu(z)$ for simulated data-set along with their residues.
$\delta h(z)$ and $\delta \mu(z)$ are the residues in the hubble parameter and distance modulus respectively.
Residues are calculated as the difference between PCA reconstruction and the corresponding $h(z)$ and 
$\mu(z)$ calculated from the cosmological constant model.    
The left column is for the  independent variable $(1-a)$, the middle   column is for $a$
and the right column is    for $z$. 
For the blue curves, there is  no reduction, that is, $M = N = 7$. The green and red curves
are obtained by the reduction of the highest and second-highest Principal Components, respectively. }
\label{fig:simu_hz_mu}
\end{figure*}

\subsection{Derived approach} 

We first reconstruct $H(z)$ and $\mu(z)$ using simulated dataset.
The reconstruction with simulated data is a check on the viability of the method.
Fig (\ref{fig:simu_hz_mu}) shows the reconstructed curves of the reduced Hubble 
parameter $h(z)$ and distance modulus $\mu(z)$ obtained 
for the simulated data.
The reconstructed curves are  for three different reconstruction variables $(1-a)$, the scale factor $a$
and the redshift $z$.  
It is clear from the plot \ref{fig:simu_hz_mu} that the PCA reconstruction produces
a consistent result when $(1-a)$ is chosen as the independent
variable.
We also plot the difference between the fiducial model and the
reconstructed curves along for a comparison.    
The plots of the residues clearly show that the reconstruction of both
$h(z)$ and $\mu(z)$ validates the reconstruction appraoch. 
The number of terms on the initial basis is fixed at $N=7$. 
We also check  our results for $N=8$ and $N=10$, and find  that the 
results do not vary significantly.
Here, we present the reconstruction curves from the variable $z$ only for the simulated data-set. Since in this case, we do not achieve a viable reconstruction, we do not analyse this further for real data.

From table \ref{Table::H_constant_fidu} we see that 
both $(1-a)$ and $a$ variables are able to predict the value of $h_0$ closer to the assumed value 
of $h_0$ in simulated data.
We calculate the error in the prediction of $h_0$ from the Covariance matrix of the PCA data-set, where we give a cut off $\chi^2$ value, $\chi^2_{cutoff}$ to mask out some patch points which lead to under-fitting \cite{Huterer2003prl,Vazirnia2021PhRvD,Huterer:1998qv}.


Further, the correlation coefficient calculation suggests that the best 
choice of number of terms for variable $(1 - a)$ in the final polynomial(\ref{polynomial_second})
is $M = N - 1$, reduction of one term from the initial polynomial expression(\ref{polynomial}).
For variable $a$ and $z$, it is $M=N$, hence there is no reduction of terms. See (\ref{appendices}).


\begin{table}
\centering
   \begin{tabular}{ | m{2cm}| m{2.5cm} |  m{2.5cm} |}
    \hline
\multicolumn{3}{|c|}{Reduced Hubble constant $h_0$} \\
\hline
\hspace{0.5cm}Variable  & \hspace{0.5cm} Simulated & \hspace{0.5cm} Observed\\
\hline
\hspace{0.5cm} $(1-a)$ & $0.674 \pm 0.118$ \vspace{0.1cm} & $0.784 \pm 0.01157$ \\
\hline
\hspace{1cm} $a$ &  $0.664 \pm 0.443$ \vspace{0.1cm} &  $0.739 \pm 0.103$ \\
\hline
  \end{tabular}
  \caption{The value of reduced Hubble constant estimated for the simulated as well as real date-set  using the derived approach. For generation of the simulated data, $h_0$ is fixed at $0.685$. 
Error associated with the estimation of reduced Hubble constant are also given.}
\label{Table::H_constant_fidu}  
\end{table}




\begin{figure*}[ht] 
\begin{center}
\includegraphics[width=9.5cm, height=5cm]{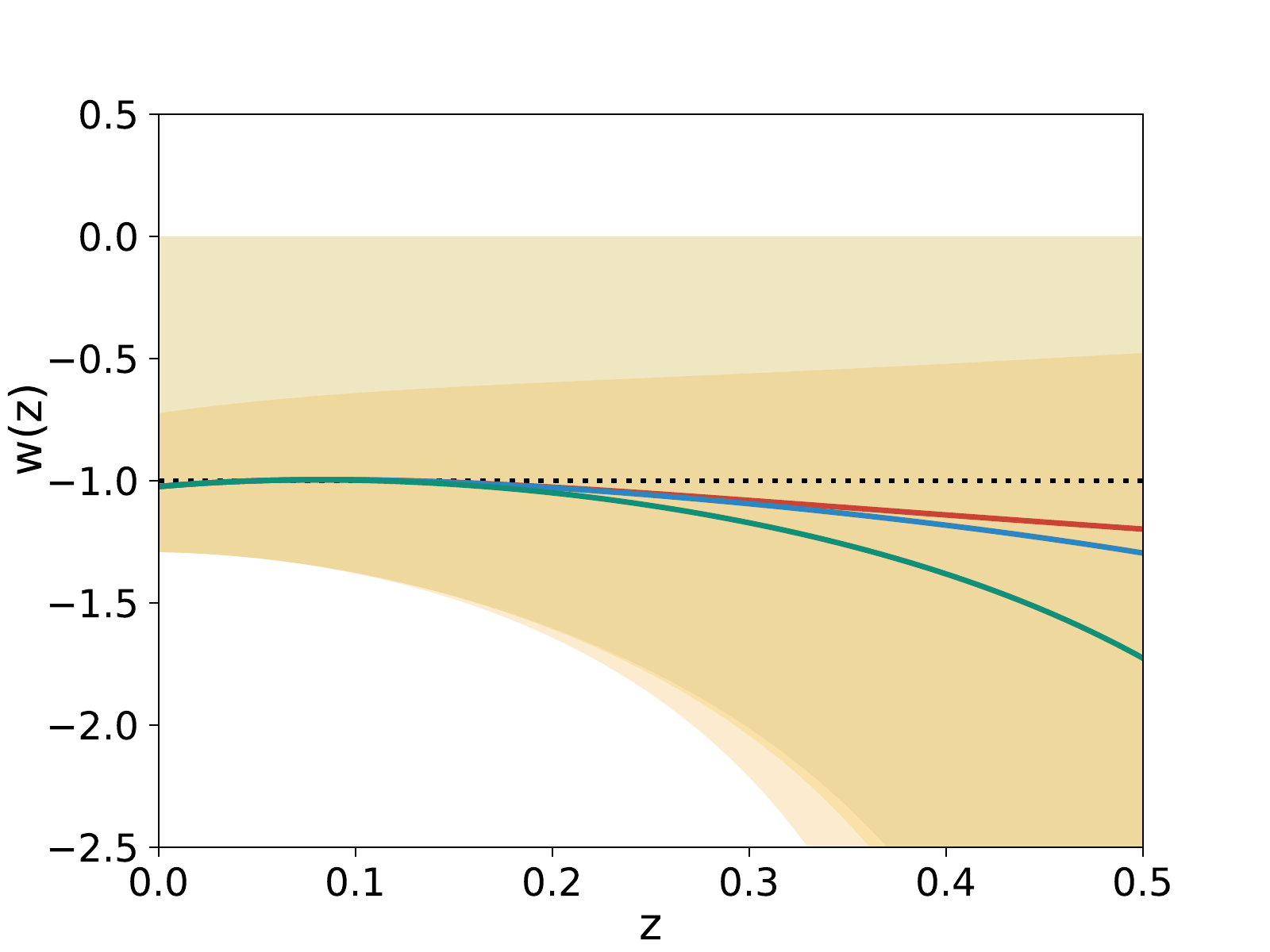}
\caption{The plots in this figure show reconstruction of $w(z)$ for the simulated data. 
Red, blue and green curves are for the reconstruction with no reduction and with 
reduction of one and two terms respectively.
In these curves, the  value of $\Omega_m$ is fixed at $0.364$, the value of $\Omega_m$ for which the reconstructed curve with no reduction and reduction of one 
term of the initial basis are the closest to the underlying $w(z) = -1$ for most of the low-redshift range.
In the figure there are three patches; grey, yellow and brown, corresponding to all the reconstruction with
$\Omega_m$ vary from 0.1 to 0.5 with no reduction of final terms and with reduction of one and two terms
respectively. 
Here we fix the reduced Hubble constant at the value predicted by PCA for the simulated data-set, $h_0 = 0.674$.
The black line is for comparison and it corresponds to $w(z) = -1$.} 
\label{fig::EoS_lcdm}
\end{center}
\end{figure*}

\begin{figure*} 
\begin{center}
\includegraphics[width=6cm, height=4cm]{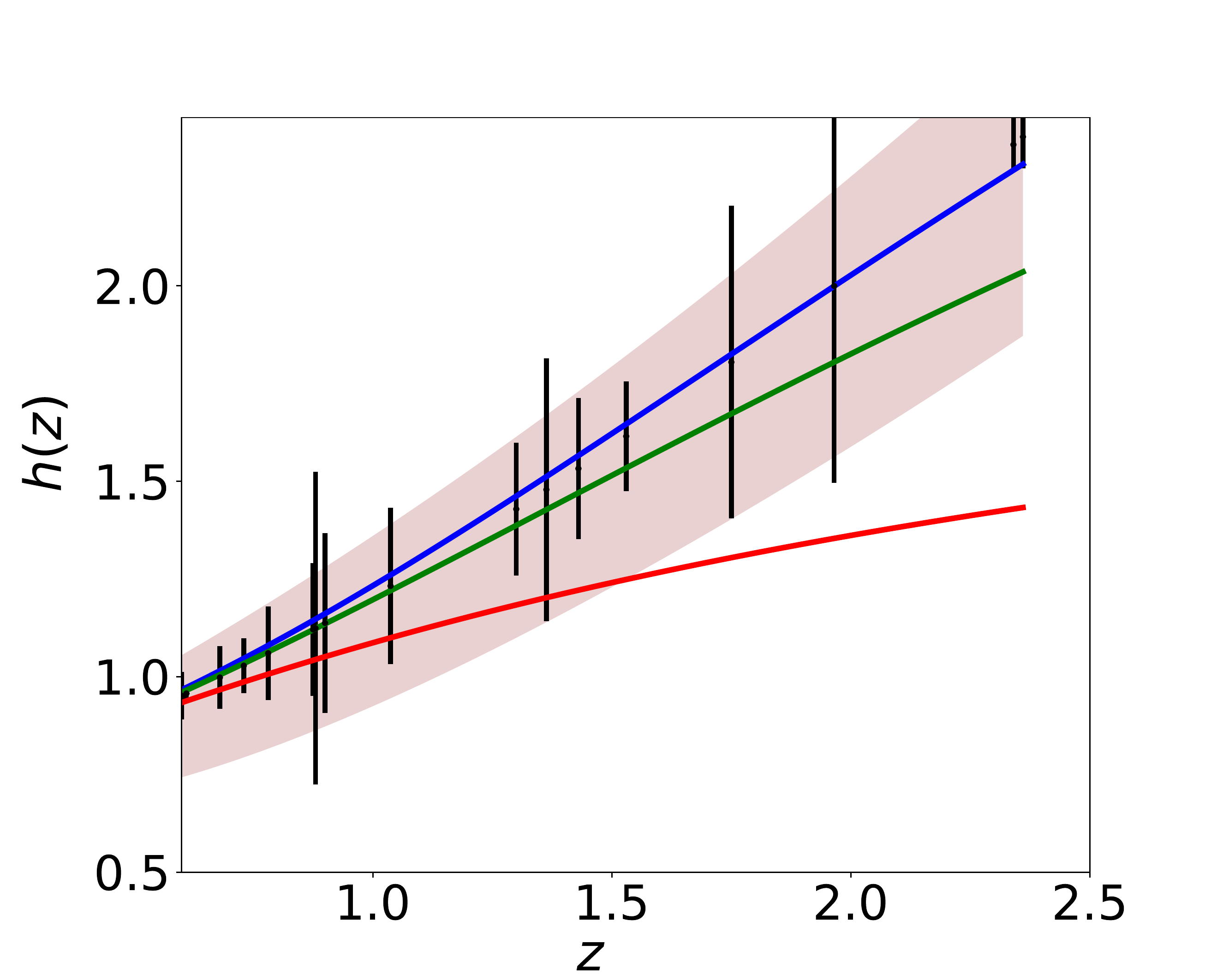}
\includegraphics[width=6cm, height=4cm]{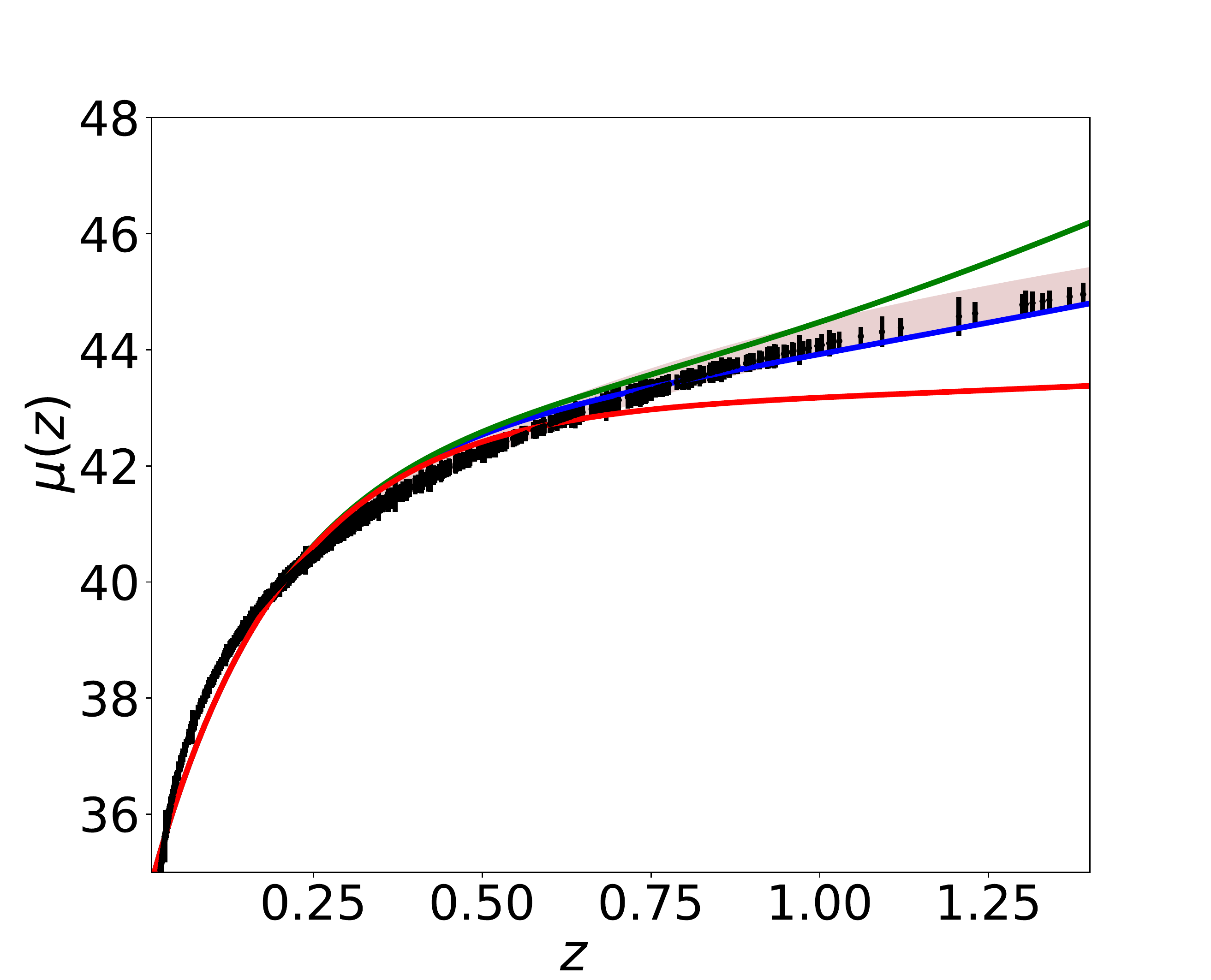}
\caption{The  plots show the reconstruction of $h(z)$ and $\mu(z)$ from the simulated 
Hubble parameter and supernovae data-set with allowed 
range $w$CDM cosmology with the initial basis function $(1 -a)$. 
The brown patch is obtained for the variation of $w$ and $\Omega_m$ in the range [-1.5 to -0.5] and [0.2 to 0.4] for $w$CDM cosmology. Solid blue, green and red line are for PCA reconstruction with no-reduction
and with one and two terms respectively.} 
\label{fig::VI_EoS_lcdm_simu}
\end{center}
\end{figure*}

\begin{figure*} 
\begin{center} 
\includegraphics[width=8cm, height=5cm]{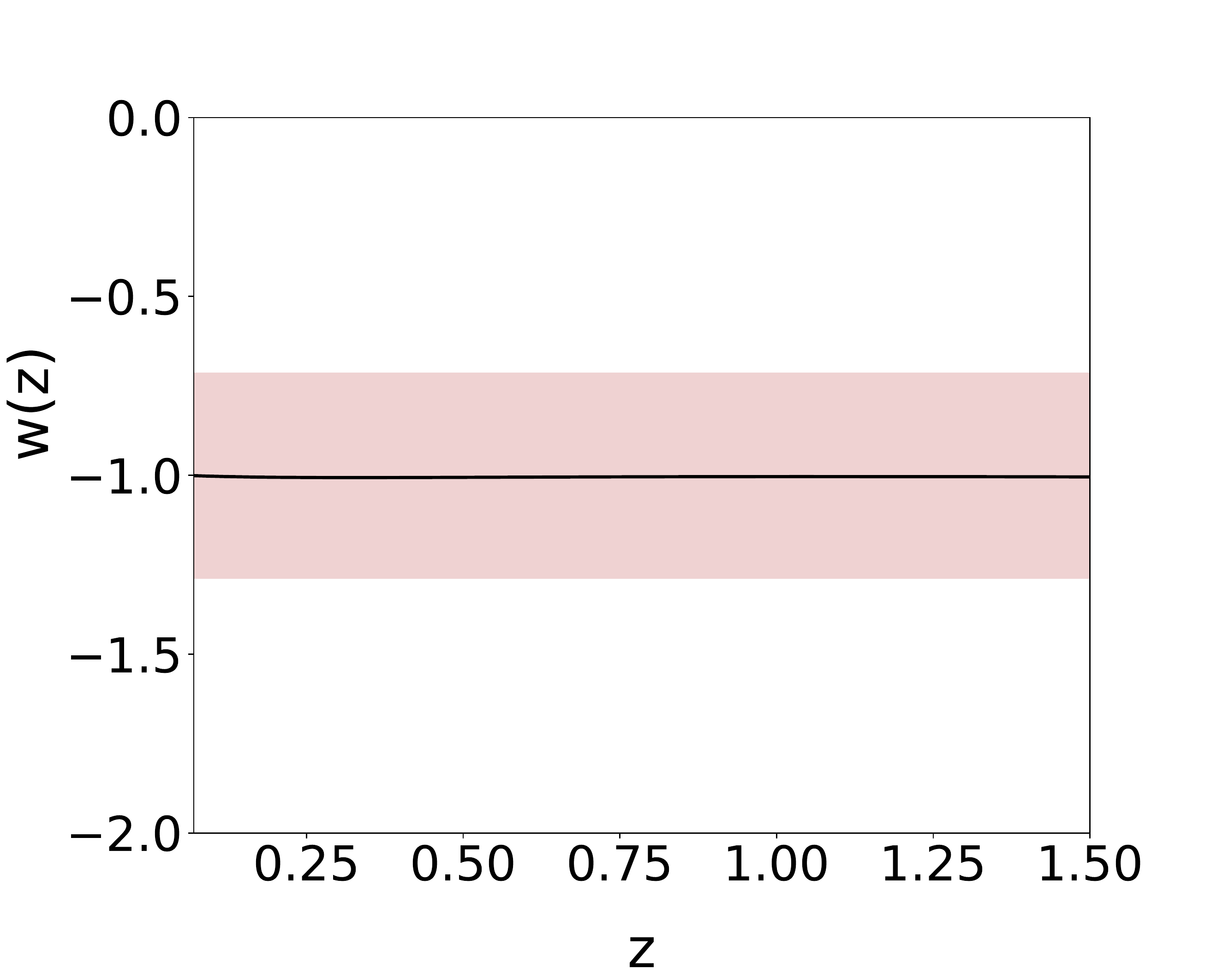}
\includegraphics[width=8cm, height=5cm]{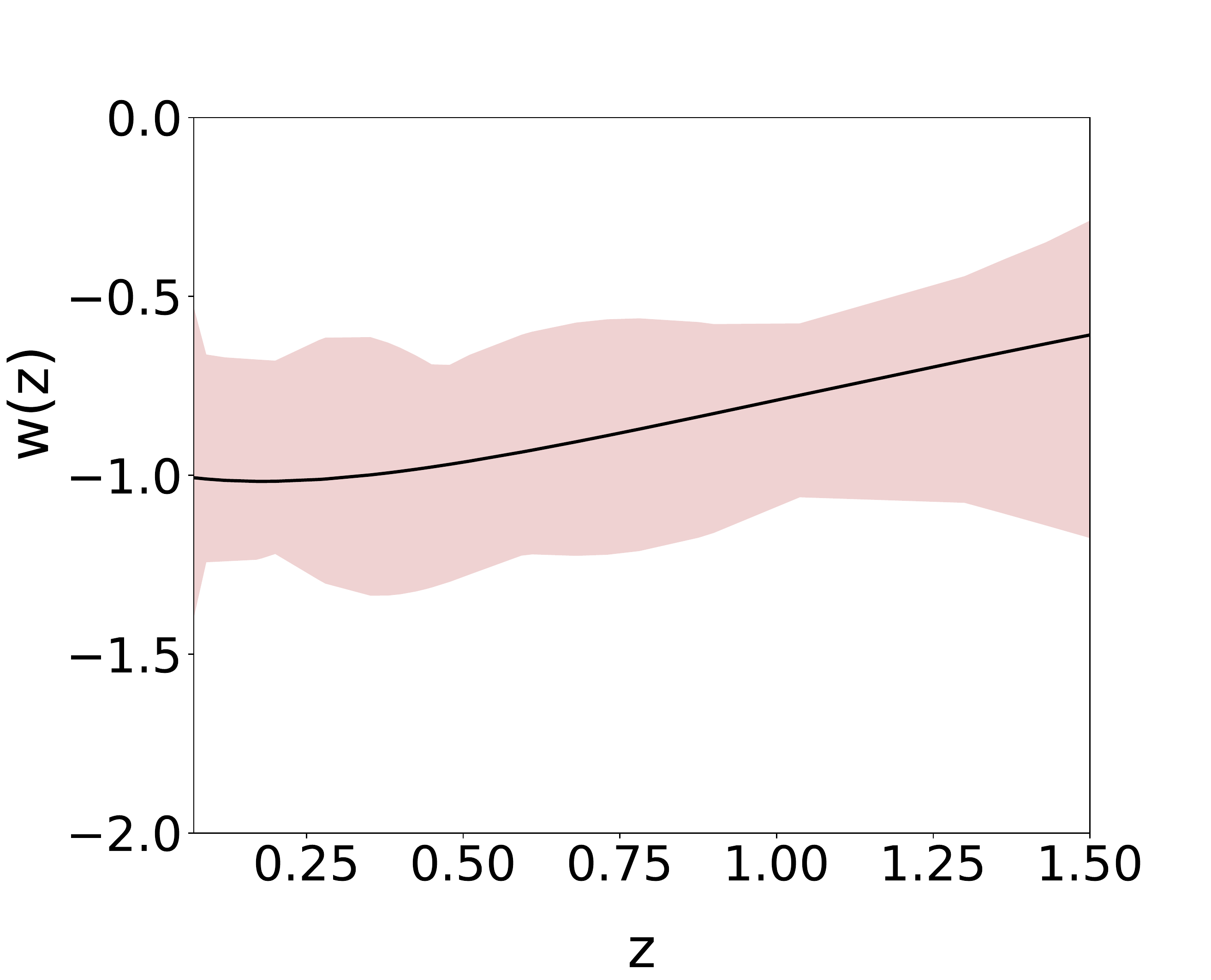}
\includegraphics[width=8cm, height=5cm]{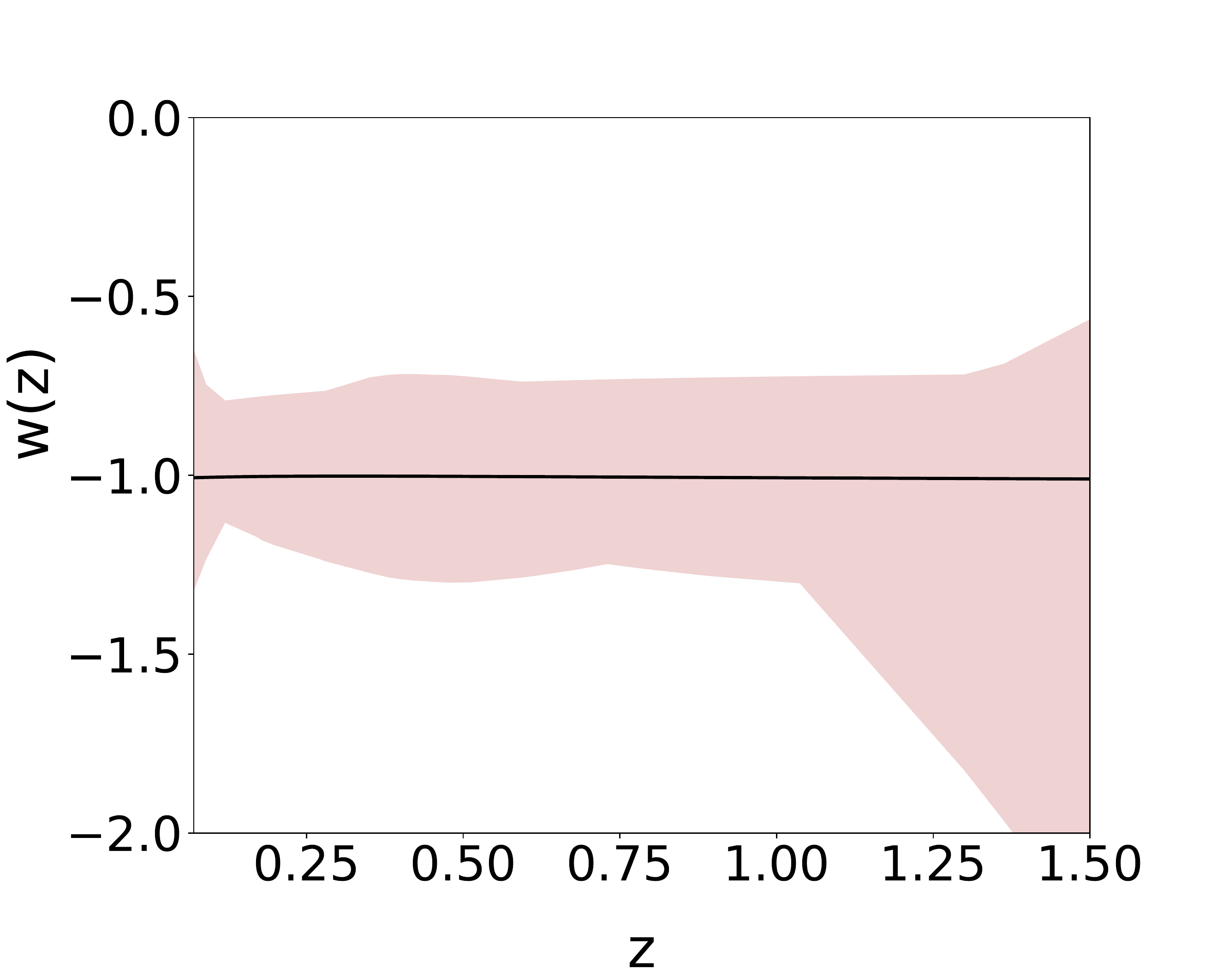}
\includegraphics[width=8cm, height=5cm]{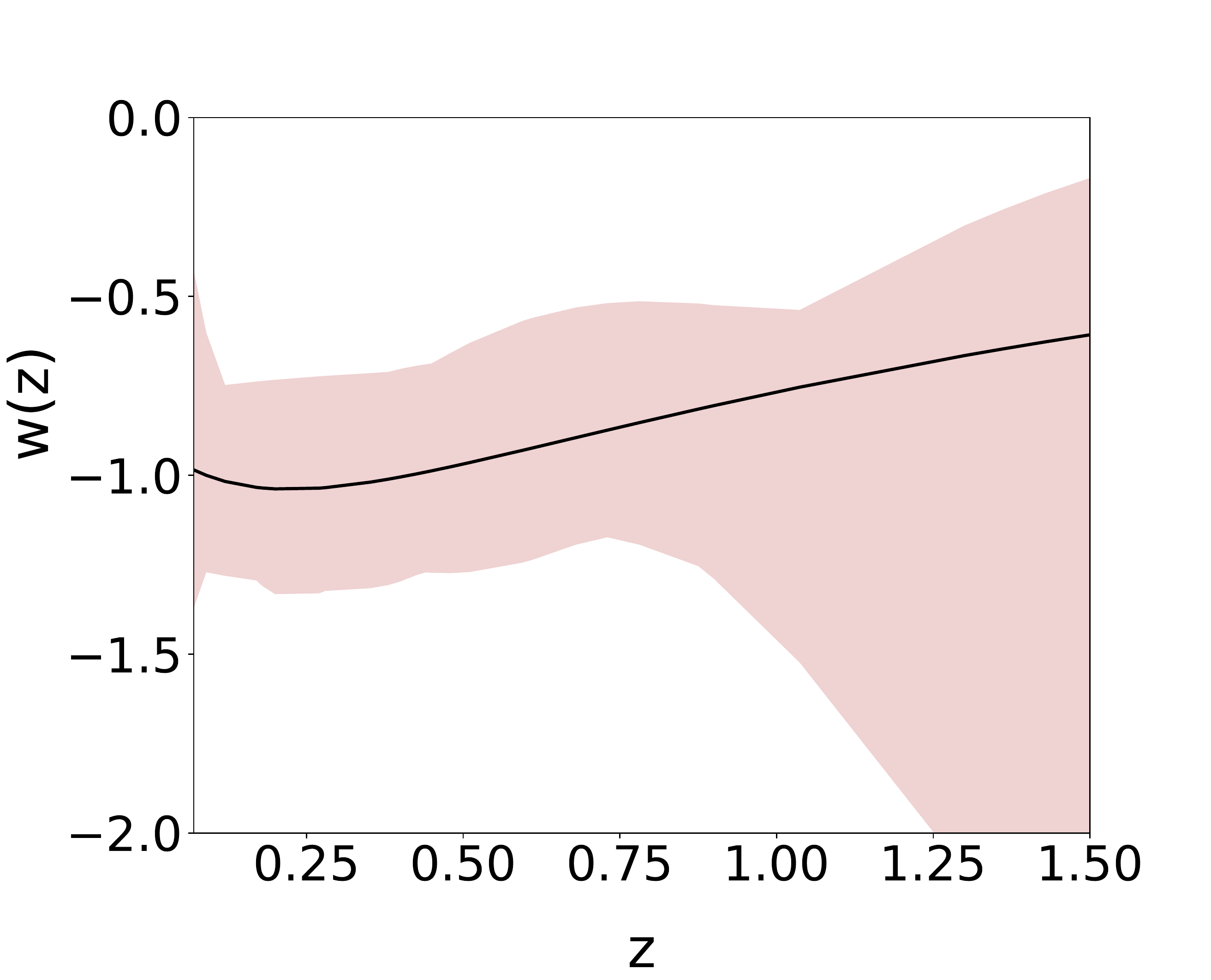}
\includegraphics[width=8cm, height=5cm]{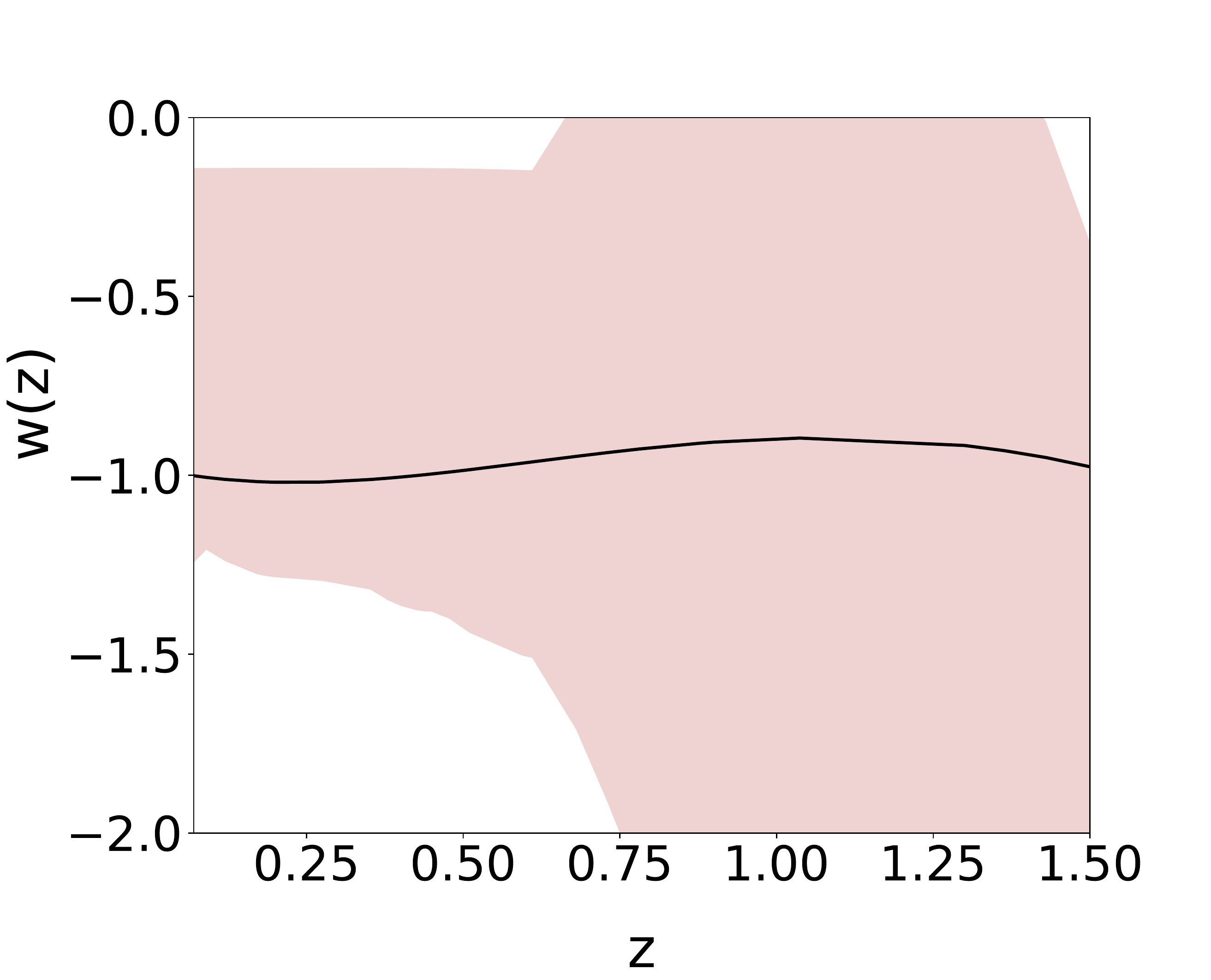}
\includegraphics[width=8cm, height=5cm]{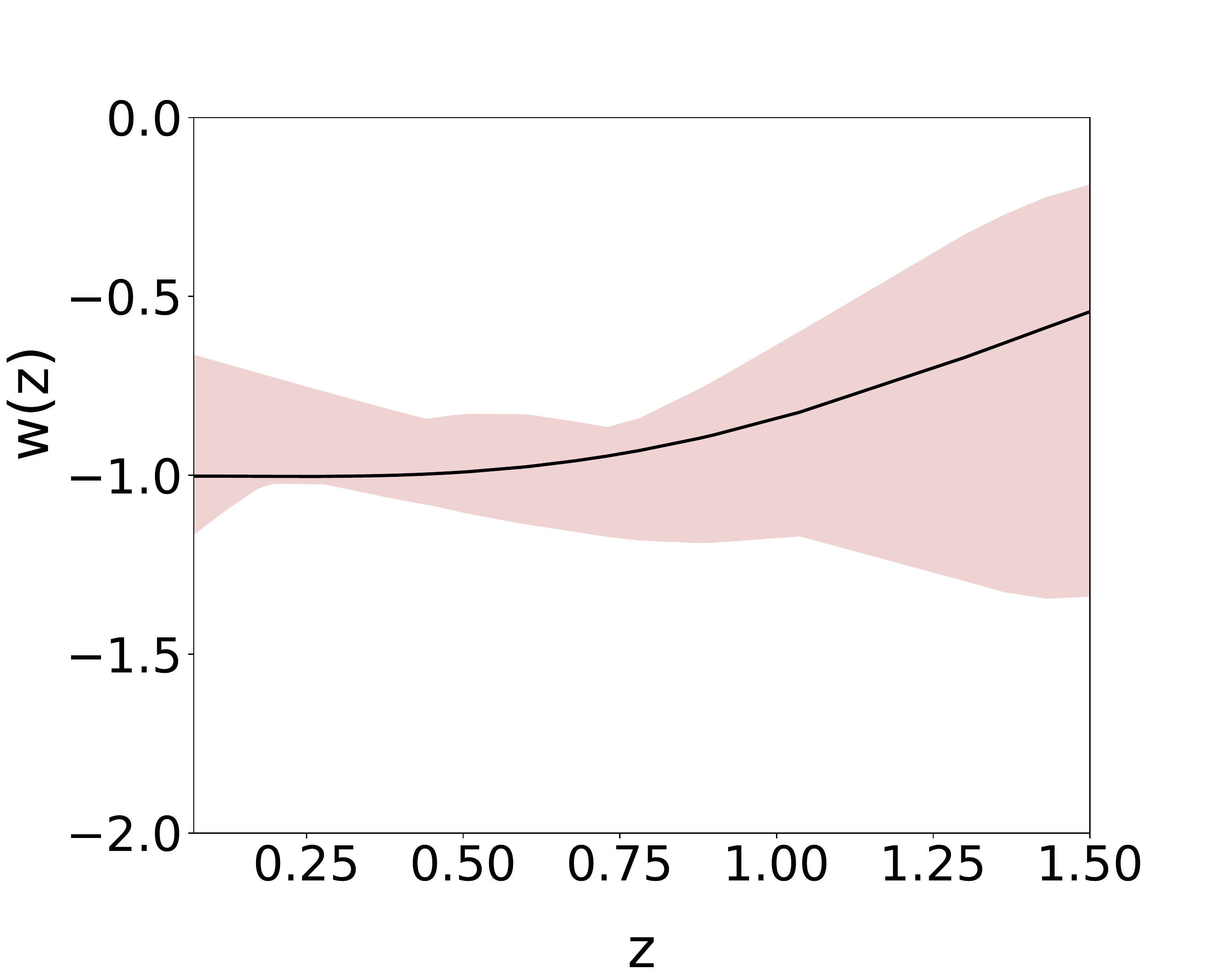}
\caption{Plots of $w(z)$ in the direct approach with simulated data-sets. The left panels are for simulated data with $w(z) = -1$ and  the right panels are for $w(z) = -\tanh{(1/z)}$. The first row is the 
reconstruction by variable $(1-a)$, second and
the third row is the reconstruction by variable $a$ and $z$, respectively. Black solid line
corresponds to  the minimum $\chi^2$ and the red shaded region
is the reconstruction using the PCA algorithm allowing a deviation of $0.3$ from the minimum $\chi^2$
curve.
} 
\label{fig:EoS_simdata_2}
\end{center}
\end{figure*}

\begin{figure*} 
\begin{center}
 \includegraphics[width=8cm, height=5cm]{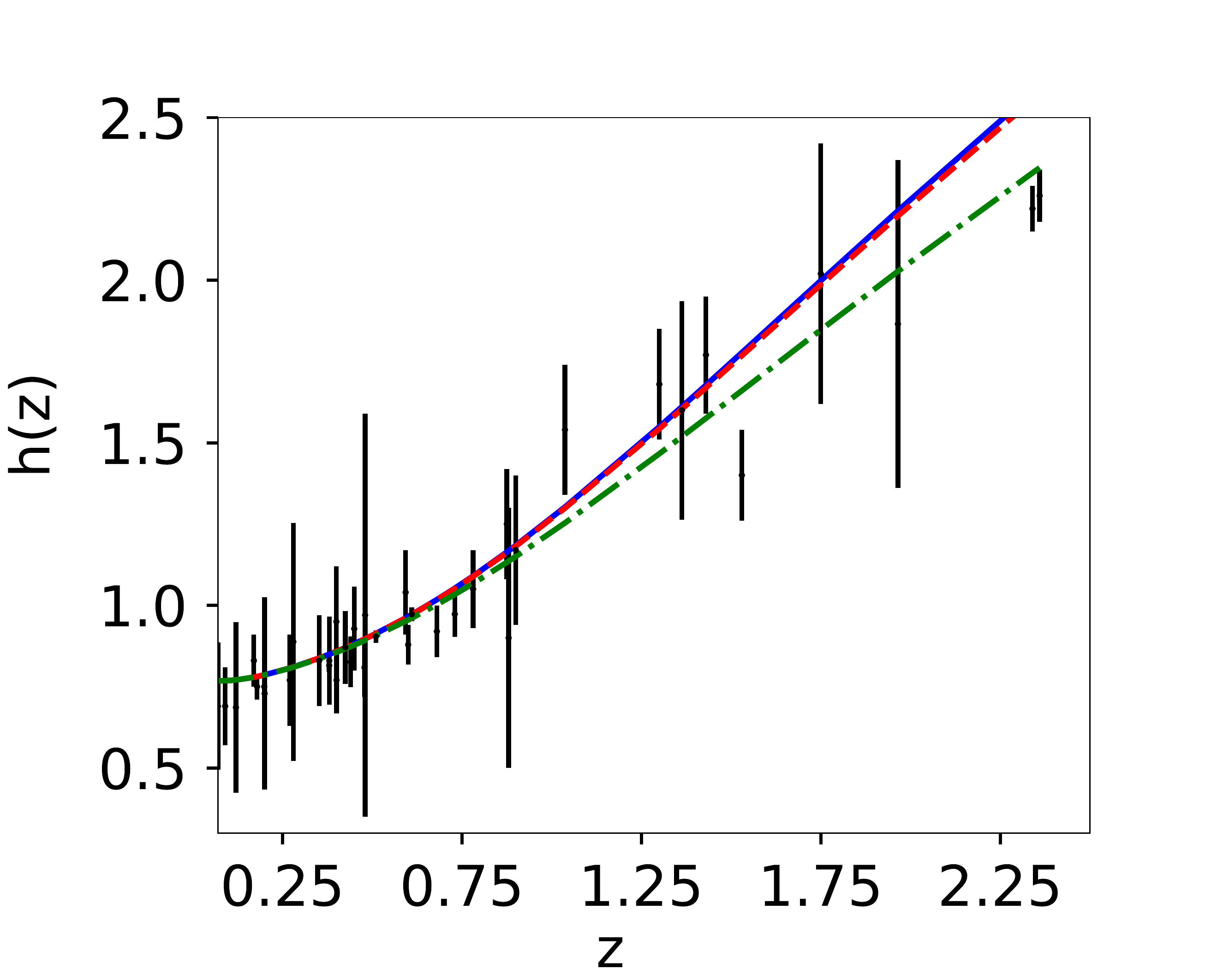}
 \includegraphics[width=8cm, height=5cm]{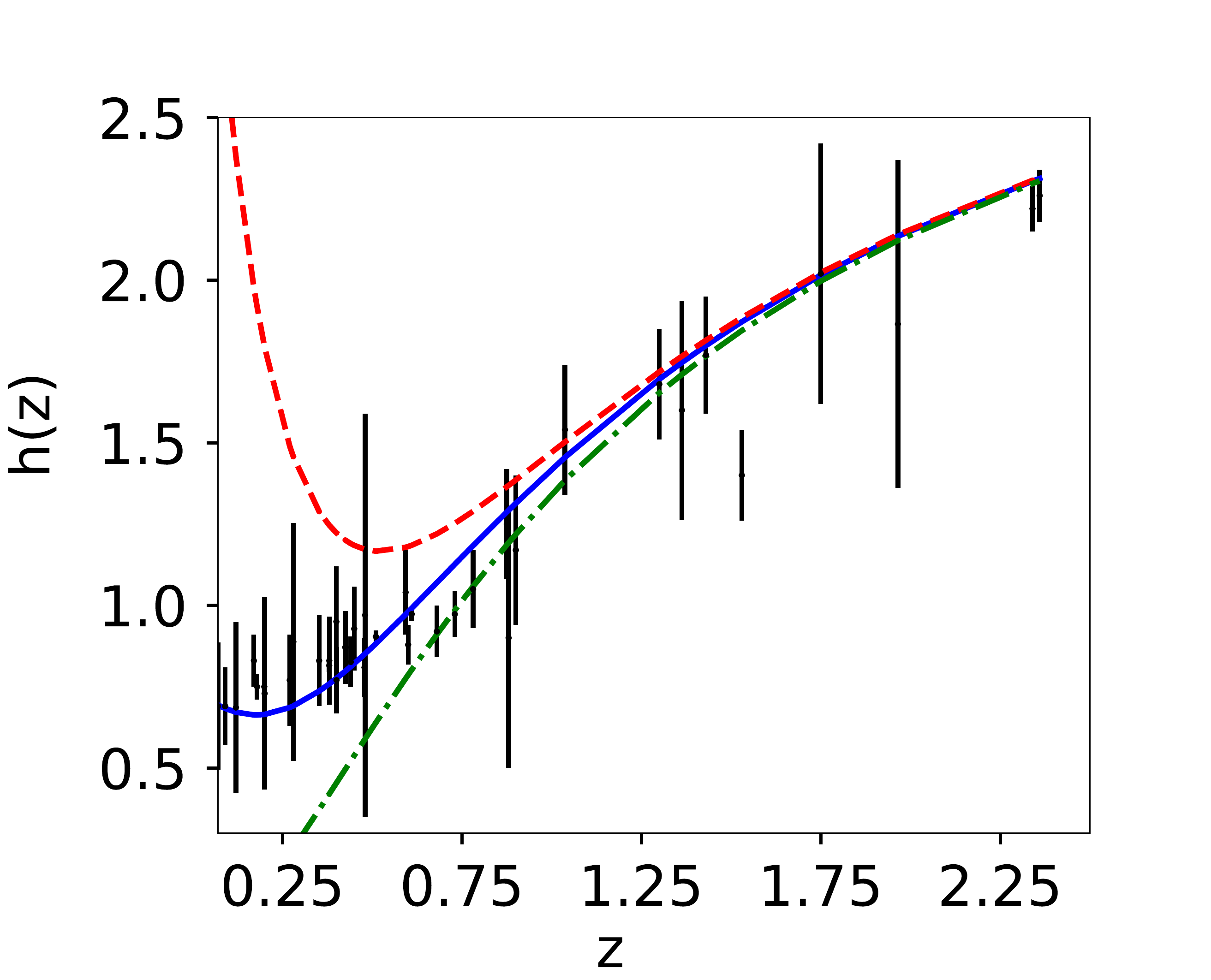} 
 \includegraphics[width=8cm, height=5cm]{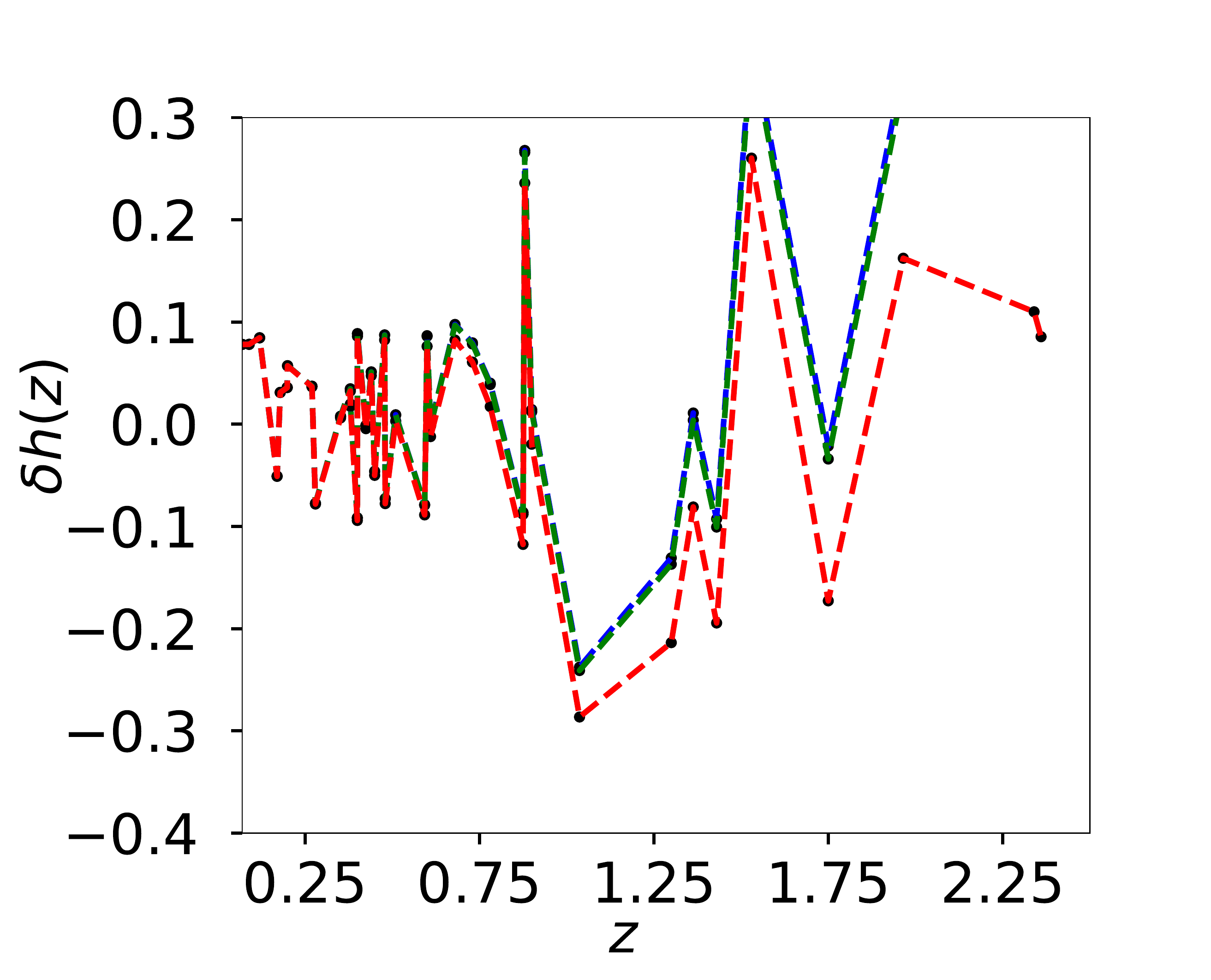}
 \includegraphics[width=8cm, height=5cm]{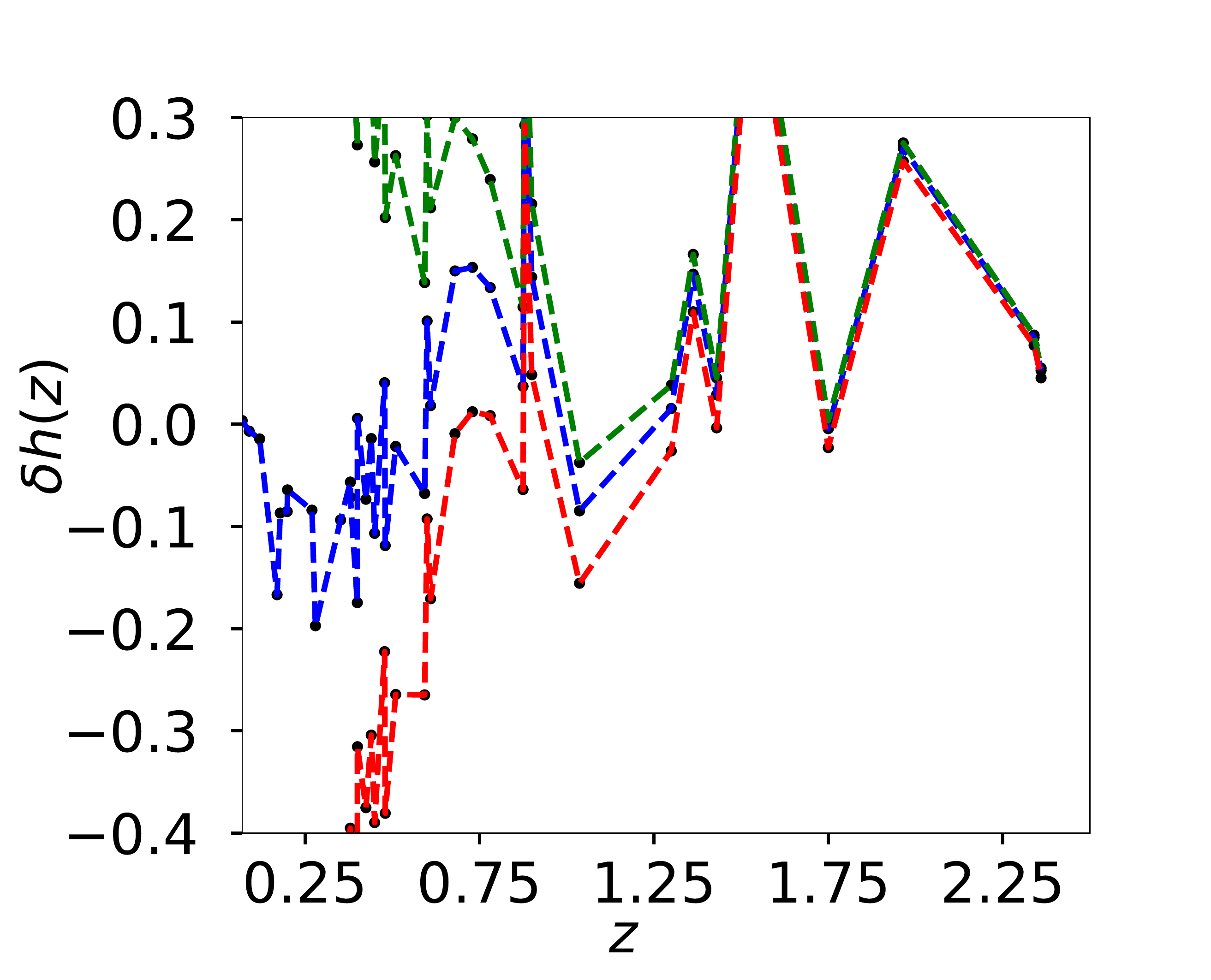}
 \includegraphics[width=8cm, height=5cm]{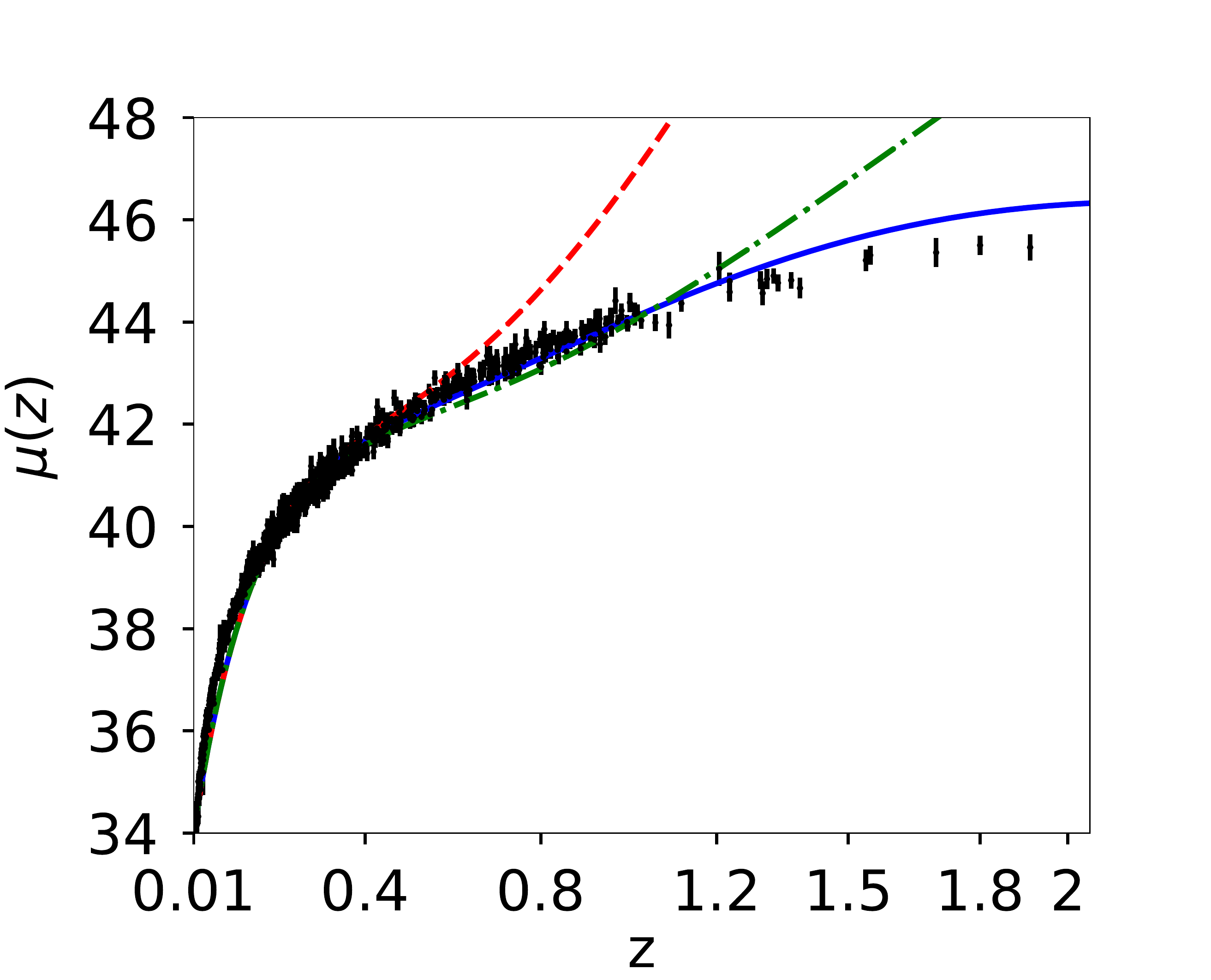}
 \includegraphics[width=8cm, height=5cm]{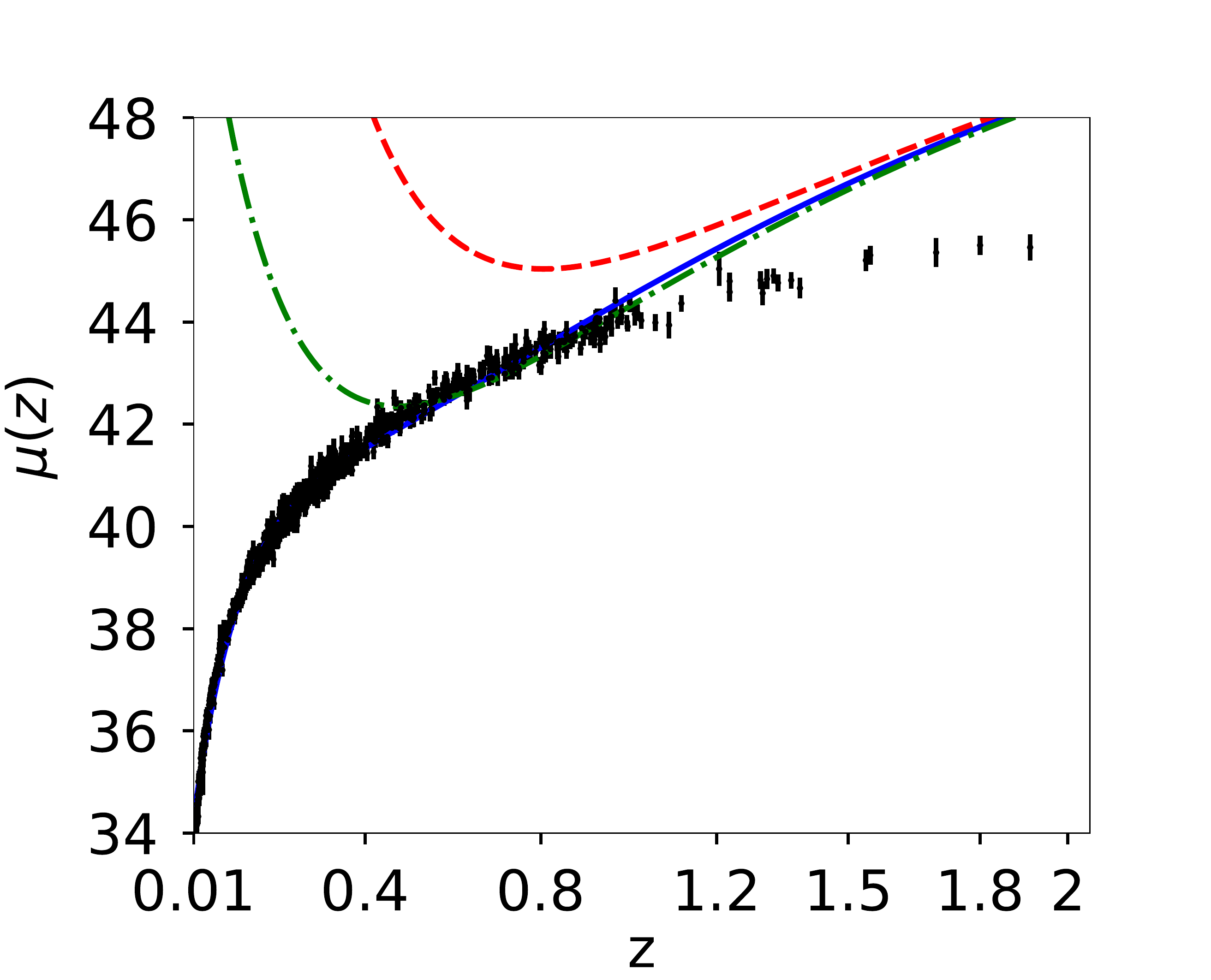}
 \includegraphics[width=8cm, height=5cm]{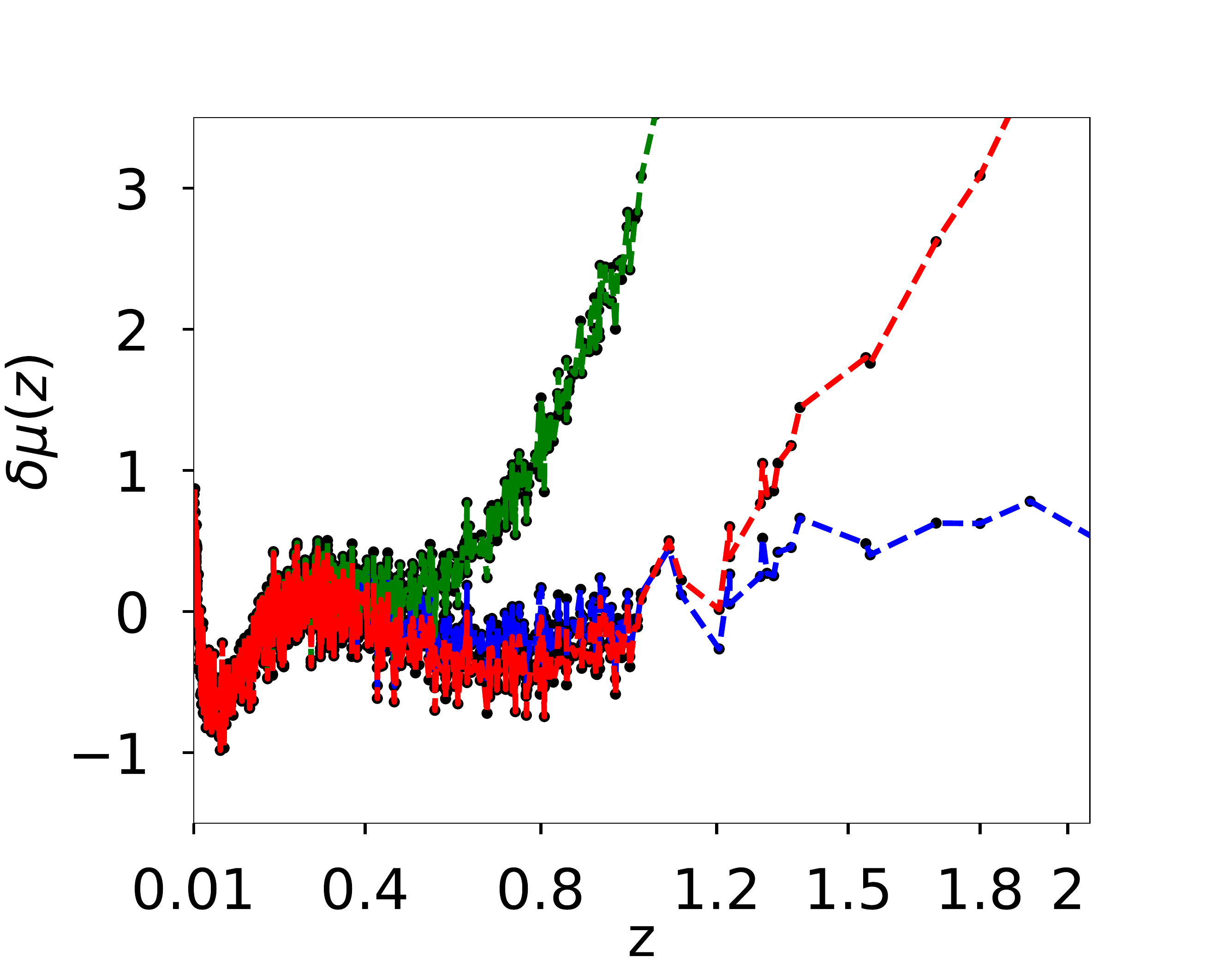}
 \includegraphics[width=8cm, height=5cm]{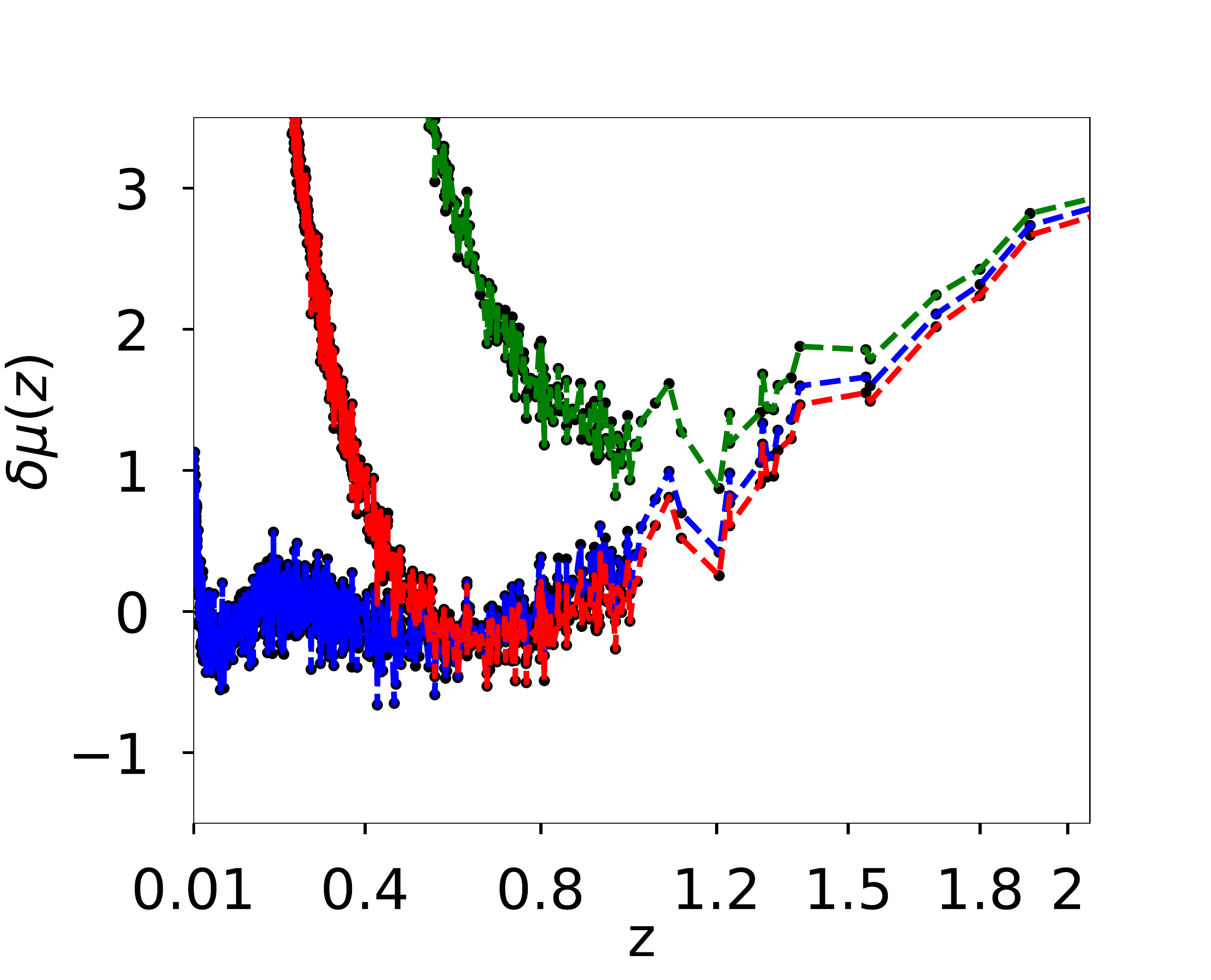}
\end{center}
\caption{The plots in the figure show the reconstructed reduced Hubble parameter $h(z)$ and
distance modulus $\mu(z)$ for observed data-sets along with their residues. 
$\delta h(z)$ and $\delta \mu(z)$ are the residue in the hubble parameter and
distance modulus respectively.
Residues are calculated as the difference between PCA reconstruction and the corresponding $h(z)$ and 
$\mu(z)$ from the observational data.    
The left column is for the  independent variable $(1-a)$, and the right column is for $a$. 
For the blue curves, there is  no reduction, that is, $M = N = 7$. The green and red curves
are obtained by the reduction of the highest and second-highest Principal Components, respectively. }
\label{fig:real_hz_mu}
\end{figure*}

\begin{figure*} 
\begin{center}
\includegraphics[width=6cm, height=4cm]{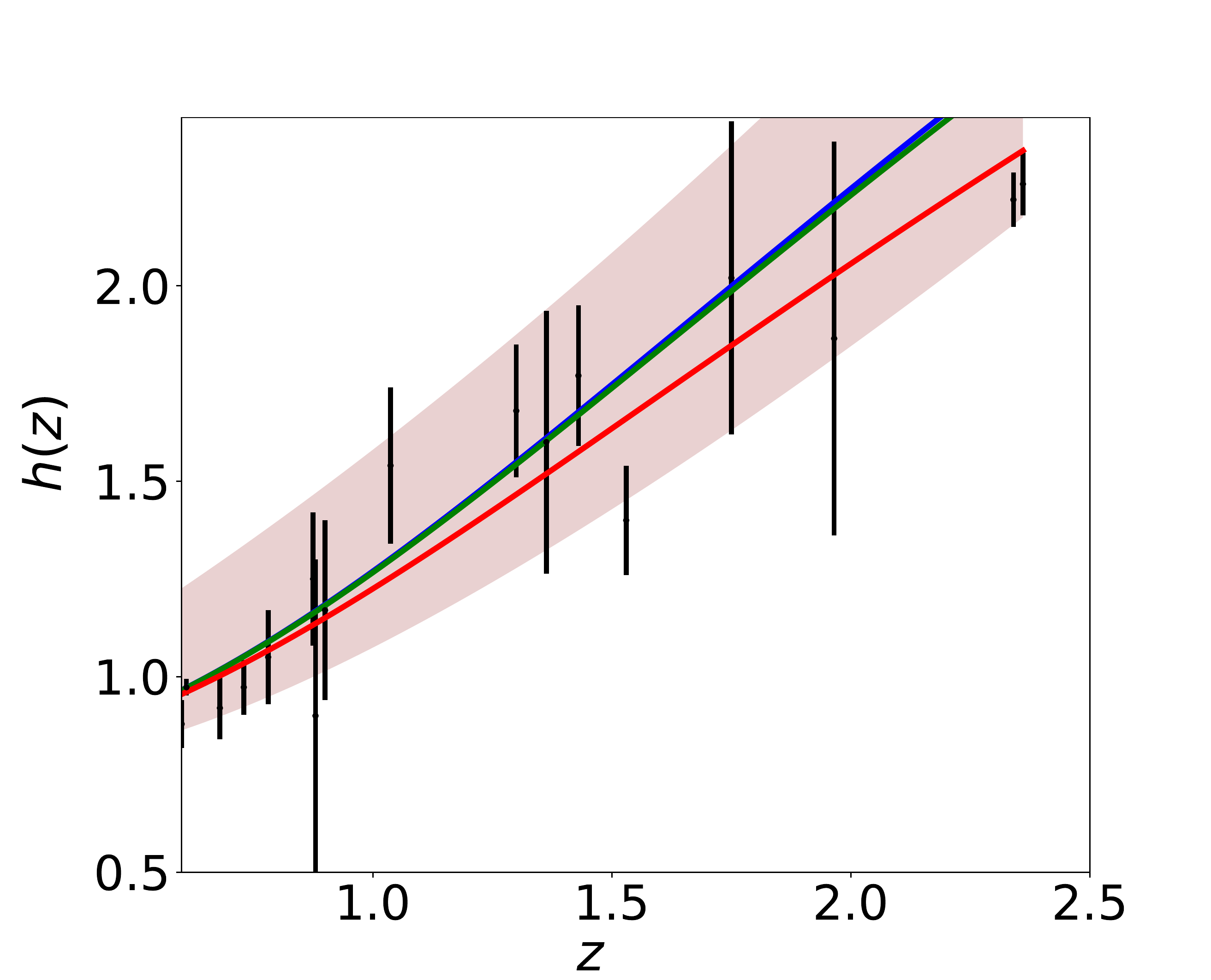}
\includegraphics[width=6cm, height=4cm]{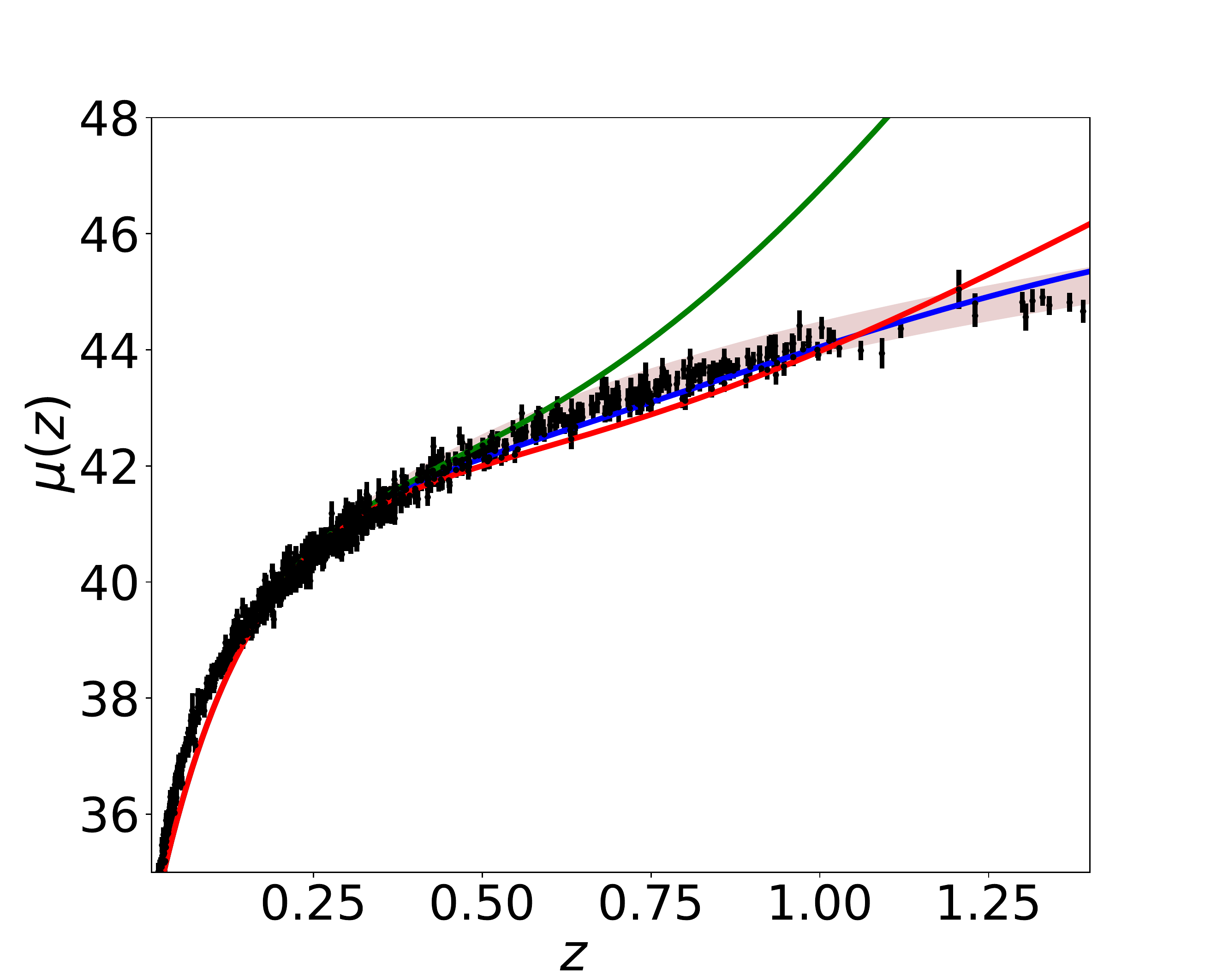}
\caption{The  plots show the reconstruction of $h(z)$ and $\mu(z)$ from the observed 
Hubble parameter and supernovae data-set with allowed 
range $w$CDM cosmology with the initial basis function $(1 -a)$. 
The brown patch is obtained for the variation of $w$ and $\Omega_m$ in the range [-1.5 to -0.5] and [0.2 to 0.4] for $w$CDM cosmology. Solid blue, green and red line are for PCA reconstruction with no-reduction
and with one and two terms respectively.} 
\label{fig::VI_EoS_lcdm_real}
\end{center}
\end{figure*}

\begin{figure*} 
\begin{center}
\includegraphics[width=6cm, height=4cm]{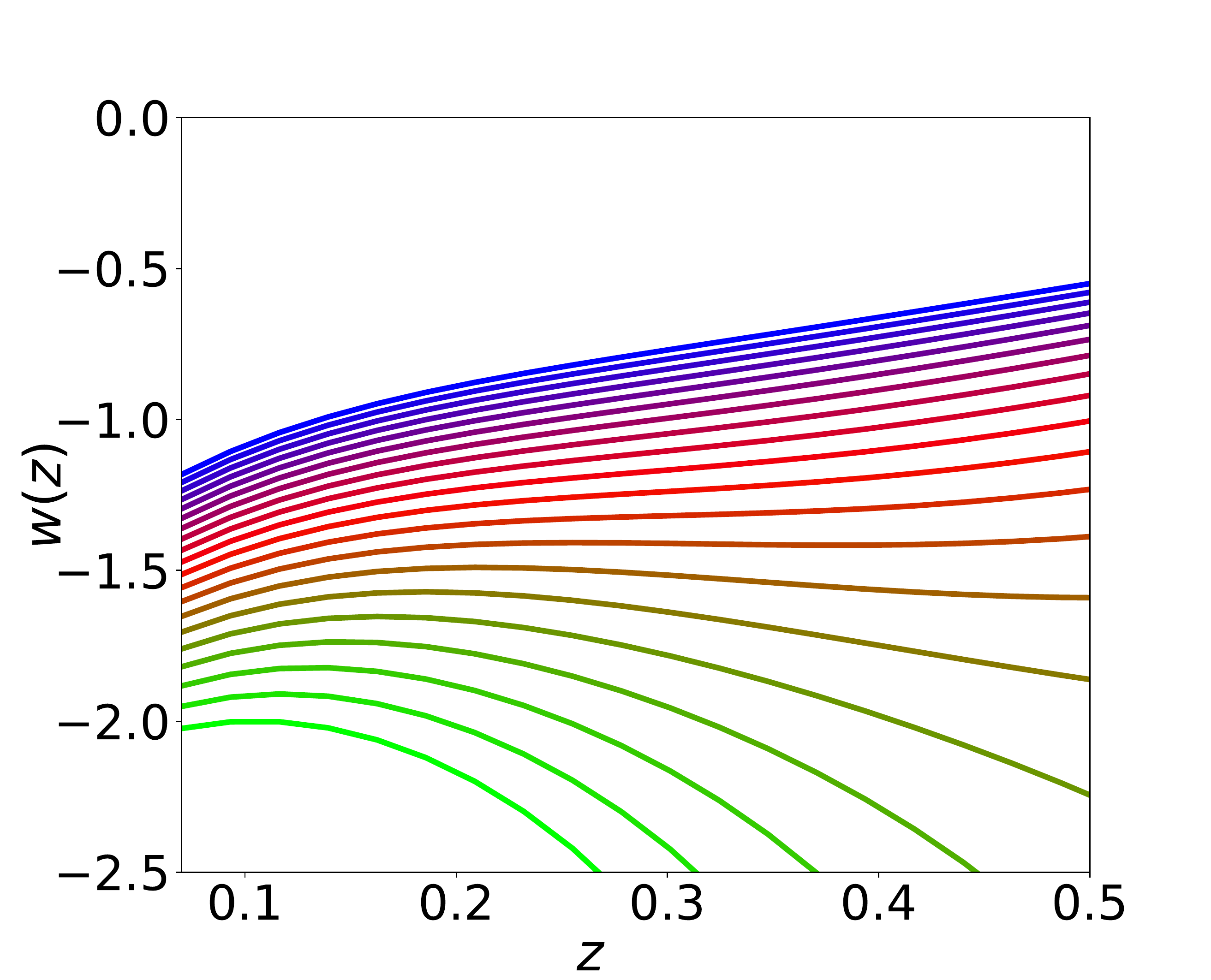}
\includegraphics[width=6cm, height=4cm]{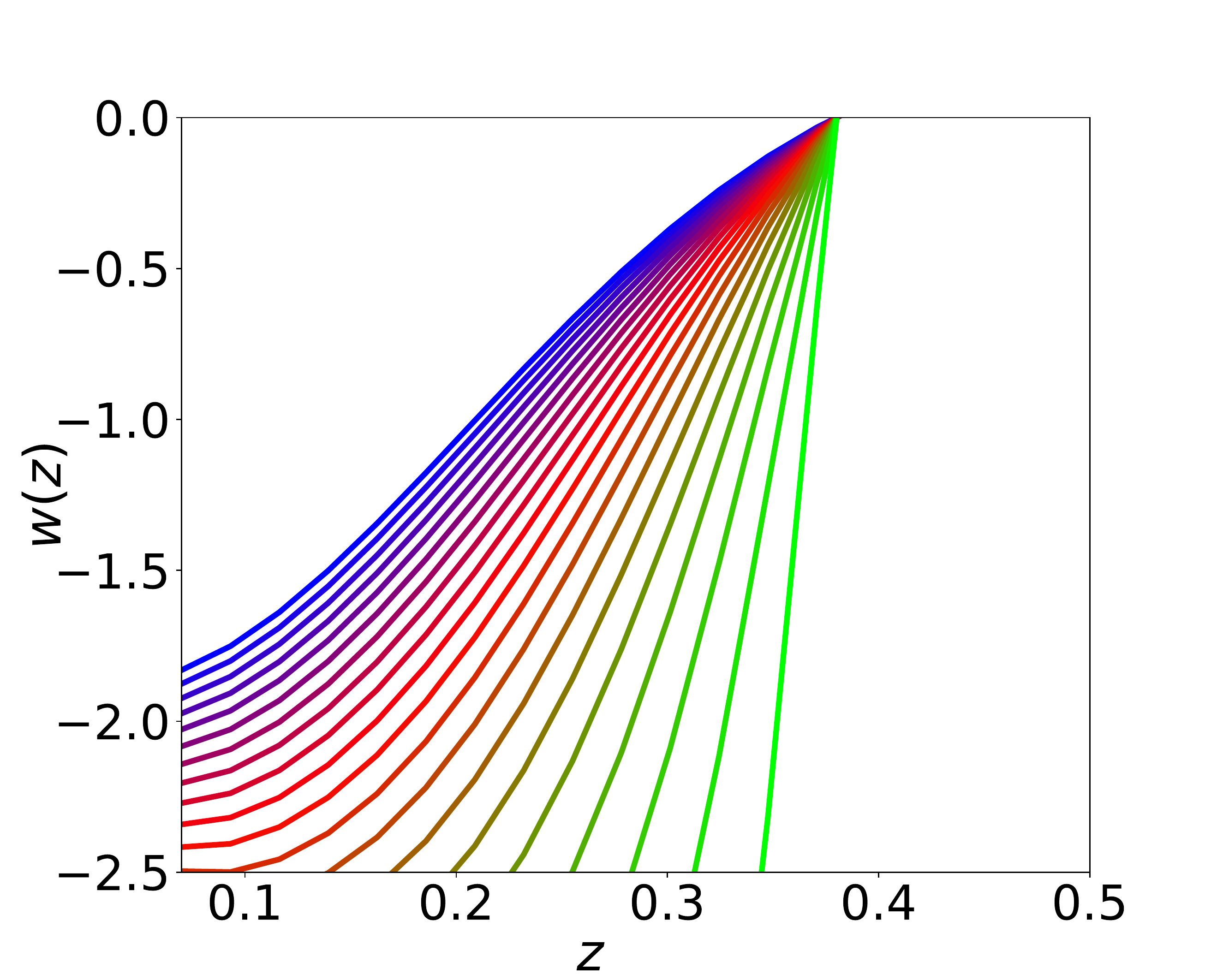}

\caption{
The figure shows  the reconstructed $w(z)$ curves, obtained in the derived approach using real Hubble parameter data, 
with variables $(1-a)$ and $a$, respectively from left.
Each represents a reconstruction with a fixed value of $\Omega_m$. 
Going from  blue to the green color variation the value of $\Omega_m$ varies from $0.1$ to $0.4$, in a step of $0.015$.
We fix the reduced Hubble constant $h_0$ at a value obtained from the PCA algorithm, as mentioned in
table \ref{Table::H_constant_fidu}.}

\label{fig:h0_constant_new}
\end{center}
\end{figure*}

\begin{figure*}
\begin{center} 
\includegraphics[width=6cm, height=4cm] {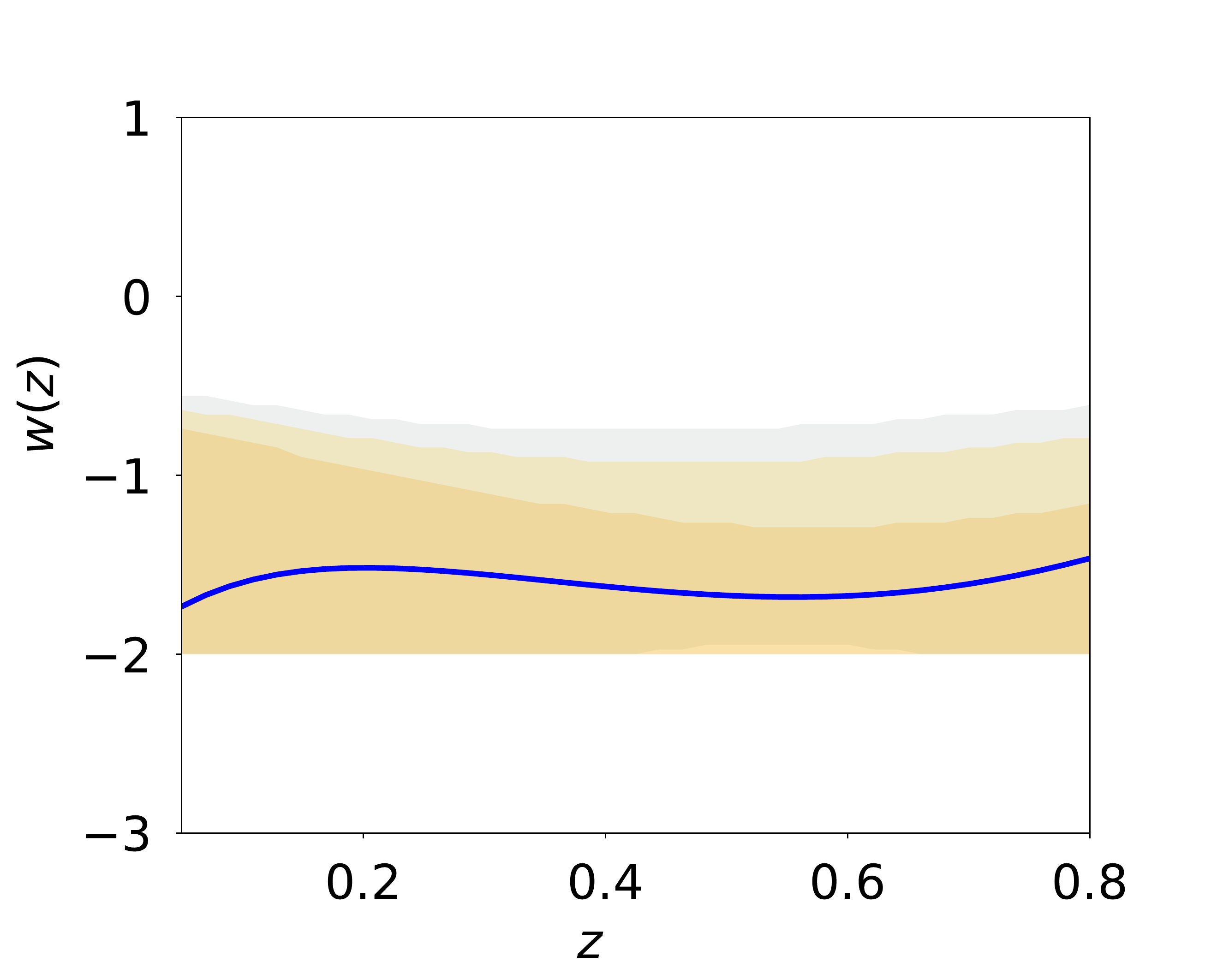}
\includegraphics[width=6cm, height=4cm]{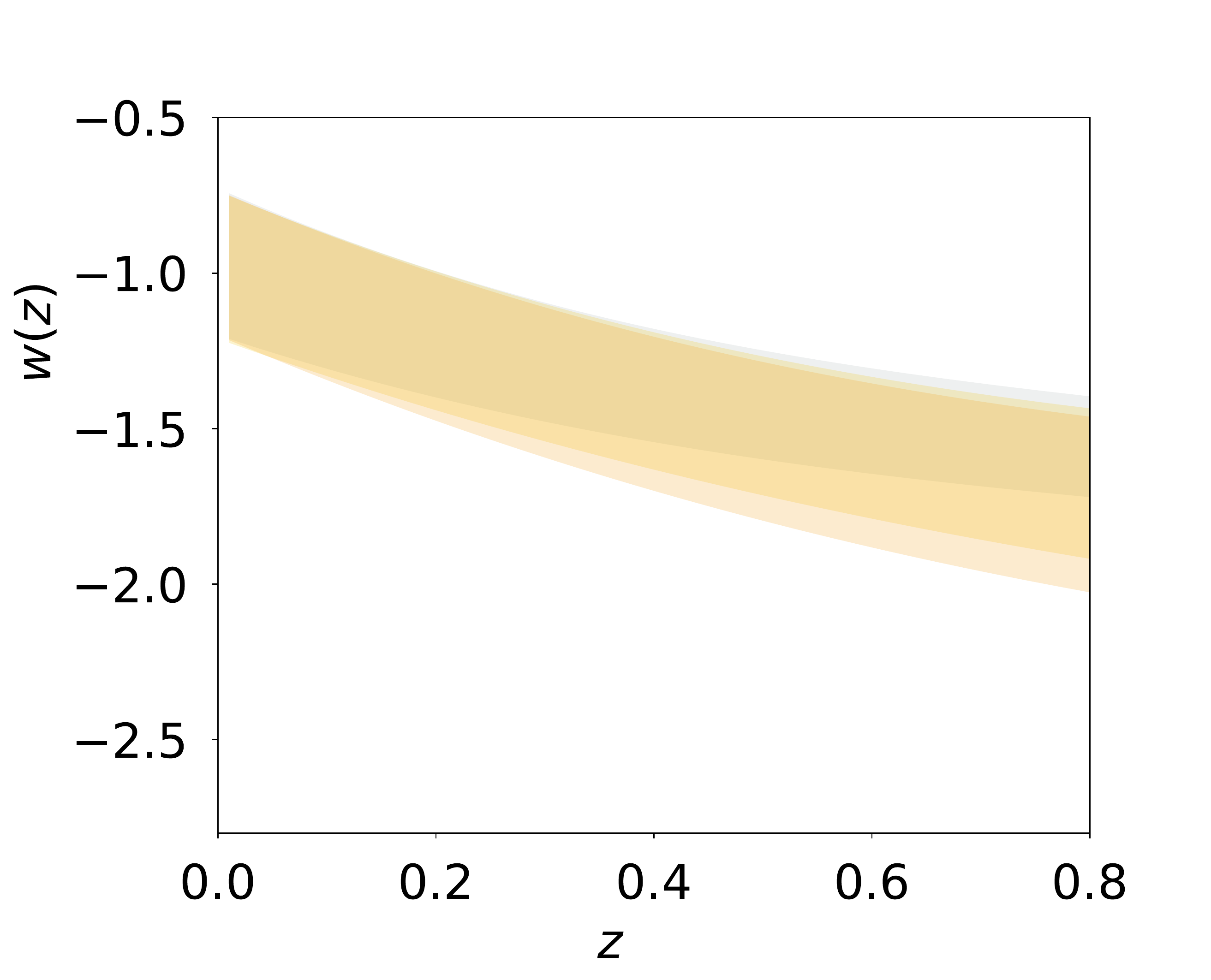}
\caption{Left and right plots are for real hubble parameter and supernovae data-set respectively. 
For both the plot Brown, yellow, grey bands are for $\Omega_m = 0.2, 0.3, 0.4$ respectively.
In the left plot blue thick line is for the $z$ vs $w(z)$ curve chosen by PCA.
Both the plots are for $(1 - a)$ and we vary $w$ in the range $[-2, 0.6]$.}
\label{fig:w_range_param}
\end{center}
\end{figure*}

\begin{table*}[t]
\centering
\begin{tabular}{|c|c|c|c|c|}
  \hline
  Variable & PCA & $\Lambda CDM$ & wCDM & Planck($\Lambda CDM$)\\
\hline
\multirow{4}{*}{(1-a)} & \multirow{4}{*}{$78.4001 \pm 1.157$} & \multirow{4}{*}{$67.94 \pm 5.15$} & \multirow{4}{*}{$68.07 \pm 1.63$} & \multirow{4}{*}{$67.9 \pm 2.6$} \\
& &\multirow{4}{*}{(Plank+WP+SDSS+SNLS)} & \multirow{4}{*}{(Plank+WP+JLA)} &  \multirow{4}{*}{(EE+lowE)}\\
& & & & \\
\multirow{4}{*}{a} & \multirow{4}{*}{$73.9271 \pm 10.3$} & \multirow{4}{*}{$69.85 \pm 4.44$} & \multirow{4}{*}{$68.19 \pm 1.33$}  & \multirow{4}{*}{$67.39 \pm 0.54$} \\
& & \multirow{4}{*}{(Plank+WP+JLA)} & \multirow{4}{*}{(WMAP9+JLA+BAO)} & \multirow{4}{*}{(TT,TE,EE+lowE+lensing)}\\
& & & & \\
& & & & \\

\hline
\end{tabular}
\caption{This is the comparison table of the values of Hubble constant in standard units ($km~s^{-1}~Mpc^{-1}$), obtained in the present analysis (PCA) and obtained from other model-dependent estimations.}
\label{Table::H_constant}
\end{table*}

Fig(\ref{fig::EoS_lcdm}) shows that
the  variable
$(1-a)$ reproduces $w=-1$ behaviour when the simulated data is used. 
Shaded region of fig(\ref{fig::EoS_lcdm}) represents all possible $w(z)$ vs $z$ curves produced by PCA for simulated data with a variation of $\Omega_m$ from $0.1$ to $0.5$. 
$\Omega_m$ is one of the free parameters in our methodology, and the methodology does not pick any one curve over another.
The polynomial expression of $H(z)$ in $(1-a)$ is effectively
an infinite series in terms of $z$. 
As the independent variable of the data-sets is $z$, therefore, $(1 - a)$ and $a$ can capture more features than the initial basis $z$.
When the observational data-set
is used in this polynomial expression, it indicates a time evolving $w(z)$, fig \ref{fig:h0_constant_new}.
The other two variables, namely $a$ and $z$, could not successfully reproduce
the $w=-1$ nature while using the simulated data. 
In the case of simulated supernovae data-set we see that all 
the $w(z)$ reconstructions for $h_0 = 0.685$ and 
$\Omega_m$ varying from $0.1$ to $0.4$ follows a  similar trend after we choose a specific basis function.
For all the three reconstruction variable $w(z)$ vs $z$ curves fluctuate between 
phantom and non-phantom regime.

We now construct the equation of state
parameter $w(z)$ from  the reconstructed $H(z)$ and $\mu(z)$.
It is clear from fig(\ref{fig::EoS_lcdm}) that the
reconstruction by derived approach for the variable $(1-a)$
successfully reproduces the $w(z)$ assumed earlier. 
On the other hand, the reconstruction variable $a$ and $z$  do not reproduce the $w(z)$
which has been assumed to simulate the data.

Using the reconstructed analytic form of $H(z)$ from derived 
approach, we estimate the present day value of the Hubble parameter $H_0$.
The EoS parameter $w(z)$ calculated by the derived approach has the free
parameter $\Omega_m$, which we vary in the reconstructed $w(z)$
curves, fig \ref{fig:h0_constant_new}.  
We present the estimated values of $H_0$,
scaled by $100~km~sec^{-1}~Mpc^{-1}$, for the analysis with simulated
and the observational data in table \ref{Table::H_constant_fidu}.
We also calculate the present day value of $w(z)$.

We calculate the range of $w(z)$ from eqn(\ref{eq:FRW}). 
We assume $w(z)$ to be constant ($w$) and find out the allowed range of $w$ for the range of $h(z)$ determined by PCA. 
This range of $w(z)$ is dependent on $\Omega_m$ and we do the analysis for $\Omega_m = 0.2, 0.3, 0.4$, as shown in fig(\ref{fig:w_range_param}).
The allowed range of $w$ for $\Omega_m = 0.2, 0.3, 0.4$ are $[-2.73, -0.71]$, $[-3.33, -0.71]$ and $[-4.14, -0.91]$  
respectively.
We do the analysis for $(1 - a)$ as it is the selected variable by correlation test calculation.
For supernovae data-set we find out the allowed range of $w(z)$, which lies within the error bar of the data-set by considering a polynomial with variable $(1 - a)$.
Range of $w(0)$ for $\Omega_m = 0.2, 0.3, 0.4$ are $[-1.2112, -0.7431]$, $[-1.224, -0.751]$ and $[-1.215, -0.7496]$ respectively.


Fig(\ref{fig::VI_EoS_lcdm_simu}) indicates that the PCA reconstructions of
  $h(z)$ is well within the range of $w$CDM
  paramters for the simulated data.
We vary $\Omega_m$ and equation of state parameter in $w CDM$
  cosmology ($w$) in  the range [$0.2,~0.4$] and [$-1.5,~-0.5$] respectively.   
The ability of our methodology to reconstruct $h(z)$ and $\mu(z)$ reflects in the reconstruction of 
$w(z)$ for the simulated data.

To quantify the efficiency of our algorithm in picking up the underlying theory, 
we create an error function by the method of interpolation with the errors of observational data 
and use $\chi^2$ with a data-set constructed from the final reconstruction curve of PCA. 
For this testing purpose we assume $w(z)$ and $\Omega_m$ as parameters, therefore
constants for a particular data-set produced from PCA reconstruction. 
We find out that PCA reconstruction for both simulated and observed data-set, if we take
$\Lambda CDM$ as our model, the value $\Omega_m = 0.3$ and $w(z) = -1$ lies well within the 
$1 \sigma$ range.

We now do the same analysis using observed data. 
Fig(\ref{fig:real_hz_mu}) shows the reconstructed curves of reduced Hubble parameter $h(z)$  and $\mu(z)$ obtained for the real Hubble parameter and supernovae data-sets.
It is evident from the plot that PCA reconstruction with $(1 - a)$ variable produces consistent results
for observational data-set also.
Fig(\ref{fig:h0_constant_new}) shows the final reconstruction of
$w(z)$ from the observed Hubble parameter data-set with the reduced
Hubble constant fixed at the value predicted by PCA. 
Reconstruction of the functional form of $h(z)$ by choosing a point
from coefficient space (eq \ref{polynomial_second}) is the first step
of derived approach, which eventually gives us the value of $h_0$ from
the same point. 
We do not vary the value of $h_0$  in the $w(z)$
reconstruction curves to ensure that both  $h(z)$ and $h_0$ are
from the same point of coefficient space. 
The values obtained from the observational data-set are  higher as
compared to the other model-dependent estimations. 
Table \ref{Table::H_constant} presents the values of $h_0$,
obtained, along with model-dependent estimations of $h_0$ from other
studies \cite{Aghanim:2018eyx,Mukherjee:2018oll,Haridasu:2018gqm} for
a comparison.  

\subsection{Direct approach}

For the direct approach, we have considered only  the Hubble parameter
data-set.   
Reconstruction is carried out for all the three independent variables
$(1-a)$, $a$ and $z$.  
As in the case of derived approach, we first use simulated data-set. 
The results obtained for the simulated data-sets are shown in fig(\ref{fig:EoS_simdata_2}). 
In this case, all three independent variables reproduce the  nature of
the equation of state parameter. 
In the direct approach, using the correlation test calculation, we
find  $N=5$ to be the best choice for the number of terms in the
initial polynomial eq(\ref{polynomial}).    
As the number of terms on initial basis is comparatively low and we
assume $M = N$ as our final basis number, the uncertainties which PCA
poses in predicting the best patch-point is relatively low. 
The tiny non-linearity we introduced in the $H(z)$ by considering a
polynomial form of  $w(z)$ will be amplified in the case of supernovae
data-set.  
Hence it makes the reconstruction much more unstable in the case of
reconstruction of $w(z)$ through distance modulus calculation.

For the observed Hubble parameter data-set too we reproduce the curves
using the same algorithm.  
In the case of direct approach, 
PCA cannot predict the value of $H_0$ and $\Omega_m$; 
therefore, these two parameters have to be fixed prior to the
analysis. 

For real data-set, all the $w(z)$ plots for the variation of $\Omega_m$ from $0.2$ to $0.4$ after fixing 
reduced Hubble constant at $h_0=0.685$ have the similar trend.
We also check the $w(z)$ plot for the variation of $h_0$
from $0.6$ to $0.8$ after fixing $\Omega_m$ at $0.30$.
Direct approach is more susceptible to model biasing as in the case of direct approach we have 
to select the value of $\Omega_m$ and $h_0$.
Correlation coefficient calculation also indicates that derived approach have more potential of breaking the correlations of the coefficients of the polynomial than the direct approach.

In the direct approach, though the reconstruction of the fiducial $w(z)$ is consistent
(fig(\ref{fig:EoS_simdata_2})), 
the correlation test calculation for
the direct approach shows that the algorithm is not able to
break the Pearson Correlation as it breaks down in 
the case of the derived approach.   
In the case of direct approach, for $(1-a)$ reconstruction, the magnitude of 
Pearson correlation coefficients decreases after applying PCA but changes signs for 
the first two principal components, whereas Kendall and Spearman correlation coefficient of $(1-a)$ 
for these two principal components assume higher negative value. 
Again, for reconstruction by the variable $a$, the Pearson Correlation
decreases, though both Spearman and  Kendall Correlation 
coefficients decrease in magnitude, it changes sign
for the first two principal components. 
For the variable $z$, up to the first two principal components, Pearson correlation
coefficients decrease, but Spearman and Kendall correlation
coefficients assume large negative value.  
From the correlation coefficient calculation 
the derived approach is selected over the direct approach and selects the 
reconstruction by the
independent variable $(1-a)$ as compared to variables $a$ and $z$.



\section{Summary and Conclusion} \label{sec::conclusion}

In this paper, we reconstruct late-time cosmology using the Principal
Component Analysis.
There are very few  prior assumptions about nature and distribution
of different components contributing to the energy of the Universe. 
Observational Hubble parameter and distance modulus
measurements of type Ia supernovae are the observable quantities that
are taken into account in the present analysis.  
We proceed in two different ways to do the reconstruction. 
The first one is a derived approach where the observable quantities
are reconstructed from the data using PCA, and then
$w(z)$ is obtained from the reconstructed quantities using
Friedman equation.  
The other approach is a direct one.  
In this case, $w(z)$ is reconstructed
directly from the observational data using PCA without any
intermediate reconstruction.  
Based on the efficiency of the method to break the correlation among the coefficients
we can select one reconstruction curve over the other.   
We achieve a better reconstruction in the derived approach as compared
to the direct approach, even though the direct approach has lesser
uncertainties than the derived approach in predicting the patch of the
best coefficient point from the $N$ dimensional coefficient space due
to the lower number of initial and final bases.  
In the  derived approach, though, we need no cosmological theory or model to determine the functional form of $H(z)$ and $\mu(z)$; however for the calculation of $w(z)$, we need to input a cosmological model. 
On the other hand, the direct approach starts with using a polynomial form of $w(z)$ in Friedman equation. Therefore, in the case of direct approach, the dependencies of $w(z)$ on the cosmological parameters are greater.

We have adopted the  simulated as well as observed data-sets
for our analysis.
Simulated data-sets are used to check the efficiency.  
For the reconstruction of $w(z)$ the analysis produces
consistent result only for the  Hubble parameter data.   
Though the reconstruction of $\mu(z)$ through the derived approach is
consistent and reconstruction of $\mu(z)$ is within the error bars of the data-set,
the result for the reconstructed $w(z)$ deviates drastically from the
physically acceptable range.    
The increase in the order of differentiation to connect $w(z)$ with
$\mu(z)$ is a possible reason for this inconsistency.   
Here we are focusing on the $w(z)$ reconstruction only from $H(z)-z$ data-set.
The reconstructed $w(z)$ by the variable $(1-a)$, obtained in
the derived approach for $H(z)-z$ data-set shows a phantom like nature, 
that is $w(z)<-1$ at present and a
non-phantom nature in the past for most of the values of $\Omega_m$.
In the case of direct approach, $w(z)$ curves show oscillations in the
phantom and non-phantom regime. 
The calculation of the correlation coefficients clearly shows a
preference for the derived approach.  
PCA lacks the efficiency of breaking the
correlation in the initial basis in case of the direct approach.  
This probably causes the inconsistency between the results obtained
in the derived and direct approaches.

The other important factor is the variable of reconstruction. 
In the present analysis, we have adopted three different
reconstruction variables, namely $(1-a)$, $a$ and $z$, in both direct
and derived approaches.  
Values of correlation coefficients after PCA select
the reconstruction variable $(1-a)$ over the other two.  
We should emphasize the result
obtained for variable $(1-a)$ by derived approach.   
Since we have a finite number of terms in the initial polynomial, due
to the constraints set by computational power, to some extent the
results depends on the assumption of polynomial expression of the
observables.  
One of our future plans is to develop an algorithm which can
inclusively select the most suitable initial basis form for a fix
computational power and a given observational data-set. 
The reconstructed curves, obtained for $(1-a)$, show that $w(z)$
shows a phantom nature at present epoch and it was in non-phantom
nature in the past.  
PCA reconstruction indicates an evolution of the dark energy equation of state parameter.

\section*{Appendices A} \label{appendices}

The correlation coefficients of the first three coefficients of the polynomial expression
eq(\ref{polynomial}) and eq(\ref{polynomial_second}) of the reconstructed quantity 
is shown in table \ref{Table::T1}.
PCA  breaks the linear correlation of the
coefficients, which is evident from table[\ref{Table::T1}]. 
Presence of the non-linear correlation in the initial coefficients complicates the process.  
As the first three terms of the ultimate expression of the
reconstructed quantity contains the dominant trend of the 
data-points, we only mention the correlation coefficients only for the first three
parameters.  
The reconstruction which can break Pearson Correlation to a  greater
extent as well as have lesser Spearman and Kendall Correlation
coefficients is selected. 
We can see from table \ref{Table::T1}
that the reconstruction by $(1-a)$, breaks the correlation to a greater
extent than in the case of $a$ and $z$.  
Variation of correlation coefficients before and after the application of PCA is similar
for both the simulated as well as the real data-set.
Difference of Pearson correlation coefficients for simulated and real data-set in case of $(1 -a)$ and $a$
is of the order of $10^{-6}$ or less. 
For the variable $z$, the difference of Pearson correlation coefficients for real and simulated
data is of the order of $10^{-4}$ or less.

\begin{table*}[h] 
\centering
  \begin{tabular}{ | m{2cm}| m{2cm}| m{3cm}| m{3cm}| m{3cm} | }
    \hline
Variable&State &Pearson & Spearman & Kendall\\
\hline
& & & &\\
\multirow{5}{*}{(1-a)} & pre-PCA & $\left[\begin{array}{ccc} 1 & -0.99 & 0.92 \\ * & 1 & -0.96 \\ * & * & 1 \end{array}\right]$ &$ \left[\begin{array}{ccc} 1 & -0.52 & 0.55\\ * & 1 & -0.94 \\ * & * &  1 \end{array}\right]$  &  $\left[\begin{array}{ccc} 1 & -0.45 & 0.52\\ * & 1 & -0.85 \\ * & * &  1 \end{array}\right]$\\
& & & &  \\
\cline{2-5}
& & & &  \\
& post PCA & $\left[ \begin{array}{ccc} 1 & 0.944 & 0.219 \\ * & 1 & 0.506 \\ * & * & 1 \end{array}\right]$ &$\left[ \begin{array}{ccc} 1 & 0.92 & 0.69\\ * & 1 & 0.63 \\ * & * &  1 \end{array}\right]$ & $\left[ \begin{array}{ccc} 1 & 0.845 & 0.69\\ * & 1 & 0.56 \\ * & * &  1 \end{array}\right]$ \\
& & & &  \\
\hline
& & & &\\
\multirow{5}{*}{a} & pre-PCA & $\left[\begin{array}{ccc} 1 & -0.98 & 0.93 \\ * & 1 & -0.98 \\ * & * & 1 \end{array}\right]$ & $ \left[\begin{array}{ccc} 1 & -0.68 & 0.60\\ * & 1 & -0.87 \\ * & * &  1 \end{array}\right]$  &  $\left[\begin{array}{ccc} 1 & -0.62 & 0.53\\ * & 1 & -0.75 \\ * & * &  1 \end{array}\right]$\\
& & & &  \\
\cline{2-5}
& & & &  \\
& post PCA 
& $\left[ \begin{array}{ccc}1 & -0.99 & 0.99\\ * & 1 & 1 \\ * & * &  1 \end{array}\right]$ & $\left[ \begin{array}{ccc} 1 & -0.30 & -0.32\\ * & 1 & -0.99 \\ * & * &  1  \end{array}\right]$ & $\left[ \begin{array}{ccc} 1 & -0.21 & -0.23\\ * & 1 & 0.968 \\ * & * &  1 \end{array}\right]$ \\
& & & &  \\
\hline
& & & &\\
\multirow{5}{*}{z} & pre-PCA & $\left[\begin{array}{ccc} 1 & -0.92 & 0.64 \\ * & 1 & -0.86 \\ * & * & 1 \end{array}\right]$ &$ \left[\begin{array}{ccc} 1 & -0.31 & 0.35\\ * & 1 & -0.72 \\ * & * &  1 \end{array}\right]$  &  $\left[\begin{array}{ccc} 1 & -0.23 & 0.35\\ * & 1 & -0.62 \\ * & * &  1 \end{array}\right]$\\
& & & &  \\
\cline{2-5}
& & & &  \\
& post PCA & $\left[ \begin{array}{ccc} 1 & 1 & -0.96 \\ * & 1 & -0.96 \\ * & * & 1 \end{array}\right]$ & $\left[ \begin{array}{ccc} 1 & 0.999 & -0.007\\ * & 1 & -0.008 \\ * & * &  1 \end{array}\right]$ & $\left[ \begin{array}{ccc} 1 & 0.998 & 0.0847\\ * & 1 & 0.0832 \\ * & * &  1 \end{array}\right]$ \\
& & & &  \\
\hline

  \end{tabular}
\caption{This table shows Pearson, Spearman and Kendall correlation coefficients between
  the coefficients of the first three terms of the series expansion of the
  reconstructed quantity for the reconstruction variables
  $(1-a)$, $a$ and $z$ respectively. This is in the derived approach for simulated Hubble parameter data. The first, third and fifth rows (pre-PCA), in the table, shows the
  correlation coefficients of the first three coefficients of the
  initial polynomial we start with, viz $b_1$,$b_2$ and $b_3$,
  eq(\ref{polynomial}) for the variables $(1-a)$, $a$ and $z$ respectively. The correlation coefficients of the
  first three coefficients of the final polynomial, viz $\beta_1$,
  $\beta_2$ and $\beta_3$, eq(\ref{polynomial_second}) given by
  the PCA algorithm is given in the second, fourth, sixth rows
  (post-PCA) for the variables $(1-a)$, $a$ and $z$ respectively. Since the correlation matrix is symmetric, here we only mention the upper diagonal terms, whereas corresponding lower diagonal terms are replaced by $*$.
}
    \label{Table::T1}
  \end{table*}
\FloatBarrier






\section*{Acknowledgements}
The authors acknowledge the use of High-Performance-Computing
facility at IISER Mohali. AM acknowledges the financial support from the
Science and Engineering Research Board (SERB), Department of Science and Technology, Government of India as a National Post-Doctoral Fellow (NPDF, File no. PDF/2018/001859).
The authors would like to thank J S Bagla for useful suggestions and discussions.
\section*{Data Availability}
The observational data-sets used in the analysis are publicly available and duly 
referred in the text. The simulated data-sets can be provided on request with appropriate justification.



\bibliographystyle{epj}
\bibliography{references}

\end{document}